\documentclass[11pt,english,a4paper,amsmath,amssymb,twocolumn,twoside]{article}
\usepackage{amsmath,amstext,amssymb,graphicx,geometry,color,subfig,babel,setspace}
\title{Non-genetic heterogeneity, criticality and cell differentiation}
\author{Mainak Pal$^{\dagger}$, Sayantari Ghosh\footnote{sayantari@gmail.com} \:and Indrani Bose\footnote{indrani@jcbose.ac.in}\\
       \small $^{\dagger}$Department of Physics, Bose Institute\normalsize \\
       \small 93/1, Acharya Prafulla Chandra Road, Kolkata-700009, India\normalsize\\
       \small $^{*}$Department of Biochemical Engineering and Biotechnology\normalsize \\
       \small Indian Institute of Technology, Delhi, Hauz Khas, New Delhi-110016, India\normalsize }
\date{} 
\begin{document}
\captionsetup[subfigure]{labelformat=empty}
\maketitle

\small 
\begin{abstract}
 The different cell types in a living organism acquire their identity through the process of cell differentiation in which the multipotent progenitor cells differentiate into distinct cell types. 
 Experimental evidence and analysis of large-scale microarray data establish the key role played by a two-gene motif in cell differentiation in a number of cell systems. The two genes express transcription 
 factors which repress each other's expression and autoactivate their own production. A number of theoretical models have recently been proposed based on the two-gene motif to provide a physical understanding 
 of how cell differentiation occurs. In this paper, we study a simple model of cell differentiation which assumes no cooperativity in the regulation of gene expression by the transcription factors. The latter 
 repress each other's activity directly through DNA binding and indirectly through the formation of heterodimers. We specifically investigate how deterministic processes combined with stochasticity contribute in bringing 
 about cell differentiation. The deterministic dynamics of our model give rise to a supercritical pitchfork bifurcation from an undifferentiated stable steady state to two differentiated stable steady states. The stochastic dynamics of 
 our model are studied using the approaches based on the Langevin equations and the linear noise approximation. The simulation results provide a new physical understanding of recent experimental observations. We further propose experimental measurements 
 of quantities like the variance and the lag-1 autocorrelation function in protein fluctuations as the early signatures of an approaching bifurcation point in the cell differentiation process.
\end{abstract}
\normalsize 
\section*{1. Introduction}
 \label{intro}
 Cell differentiation is a key biological process in which the stem or progenitor cells diversify into different cell types or lineages 
 in response to internal cues or external signals. Stem cells exist in both the embryo and the adult tissues. In mammals, the development of the 
 different cell types from the embryonic stem cells can be visualized by invoking the metaphors of Waddington's epigenetic landscape and the cell-fate 
 tree \cite{Zhou,MacArthur,Enver,Furusawa,Ferrel}. The epigenetic landscape consists of a cascade of branching valleys starting with a single 
 valley at the top. The parent valley bifurcates into two new valleys separated by a ridge. The valleys undergo a cascade of further 
 splits till a set of terminal valleys is reached. The topmost valley represents the undifferentiated, say, the embryonic cell state 
 whereas the terminal valleys depict the distinct cell types the stability of which is maintained by the high ridges separating the valleys. 
 The progression from the top to the bottom captures the successive stages of development. The cell fate tree \cite{Zhou} consists of branches 
 and sub-branches representing the developmental paths of different cell types. At each branch point, a cell makes a choice between two 
 prospective lineages. The branch points thus correspond to the pluripotent or progenitor cells which are still uncommitted. The terminally 
 different cells are represented by the terminal branches or leaves of the tree. The adult stem cells, appearing at the later stages of the cell-fate tree, are 
 present in small numbers in the tissues such as the skin and the bone marrow and serve to replace only the tissue-specific cell types which are lost due 
 to damage or death. Their differentiation potential is thus more limited than that of the embryonic stem cells.
 
 The distinct cell types, in general, have identical genetic content but are distinguished by different gene expression profiles. A gene 
 expressed in one cell type may be silent in another type or have a considerably modified expression level. A cell is a dynamical system 
 the state of which, represented by the relevant gene expression levels, say, the protein concentrations, changes as a function of time. 
 The time evolution is controlled by the regulatory interactions between the different genes and the external
   \begin{figure}[h]
   \centering
   \subfloat[( i )\:\:\:]{%
     \includegraphics[scale=0.4,width=0.25\textwidth]{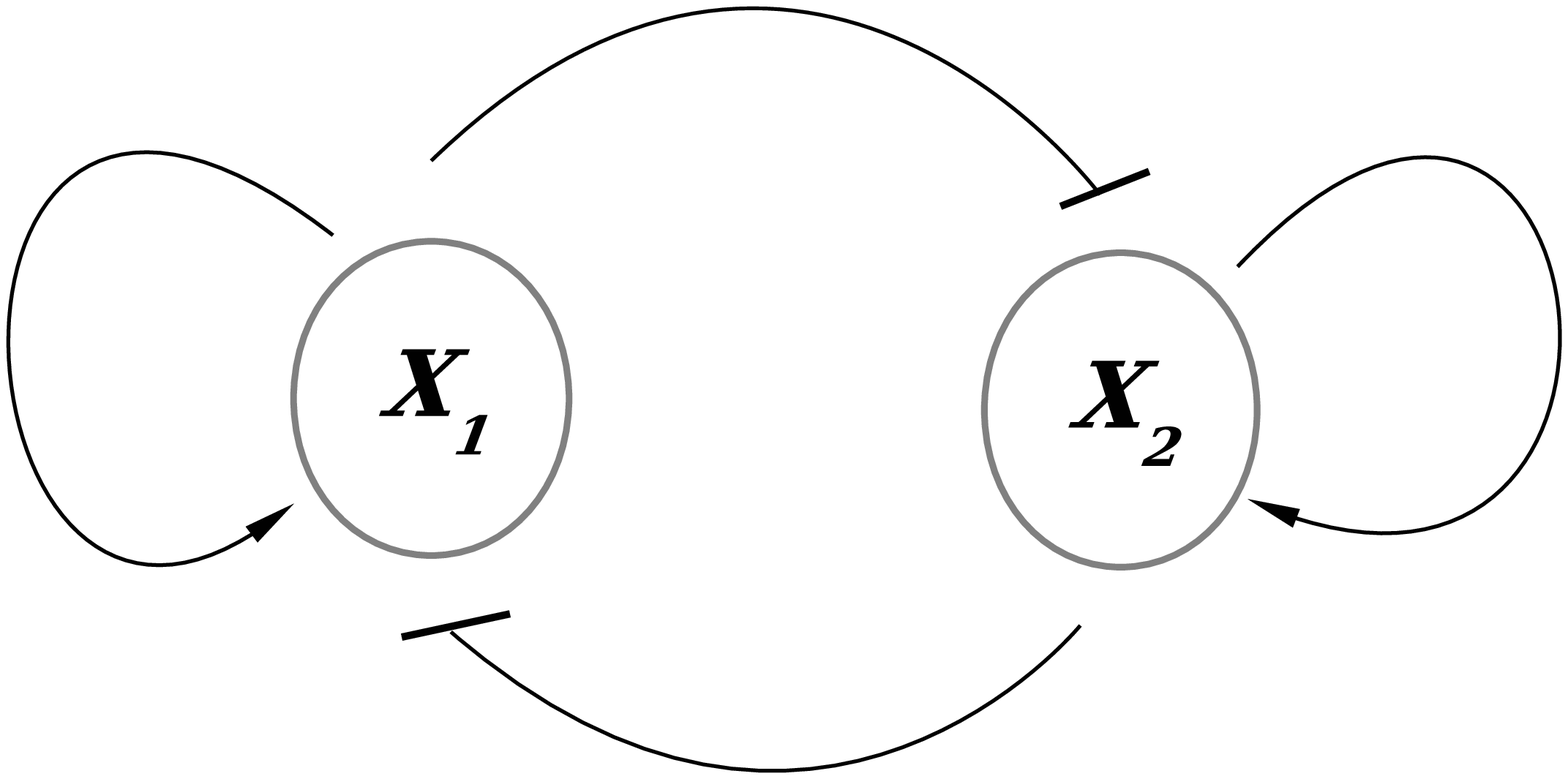}}
   \quad
   \subfloat[\:( ii )\:\:\:]{%
     \includegraphics[scale=0.4,width=0.25\textwidth]{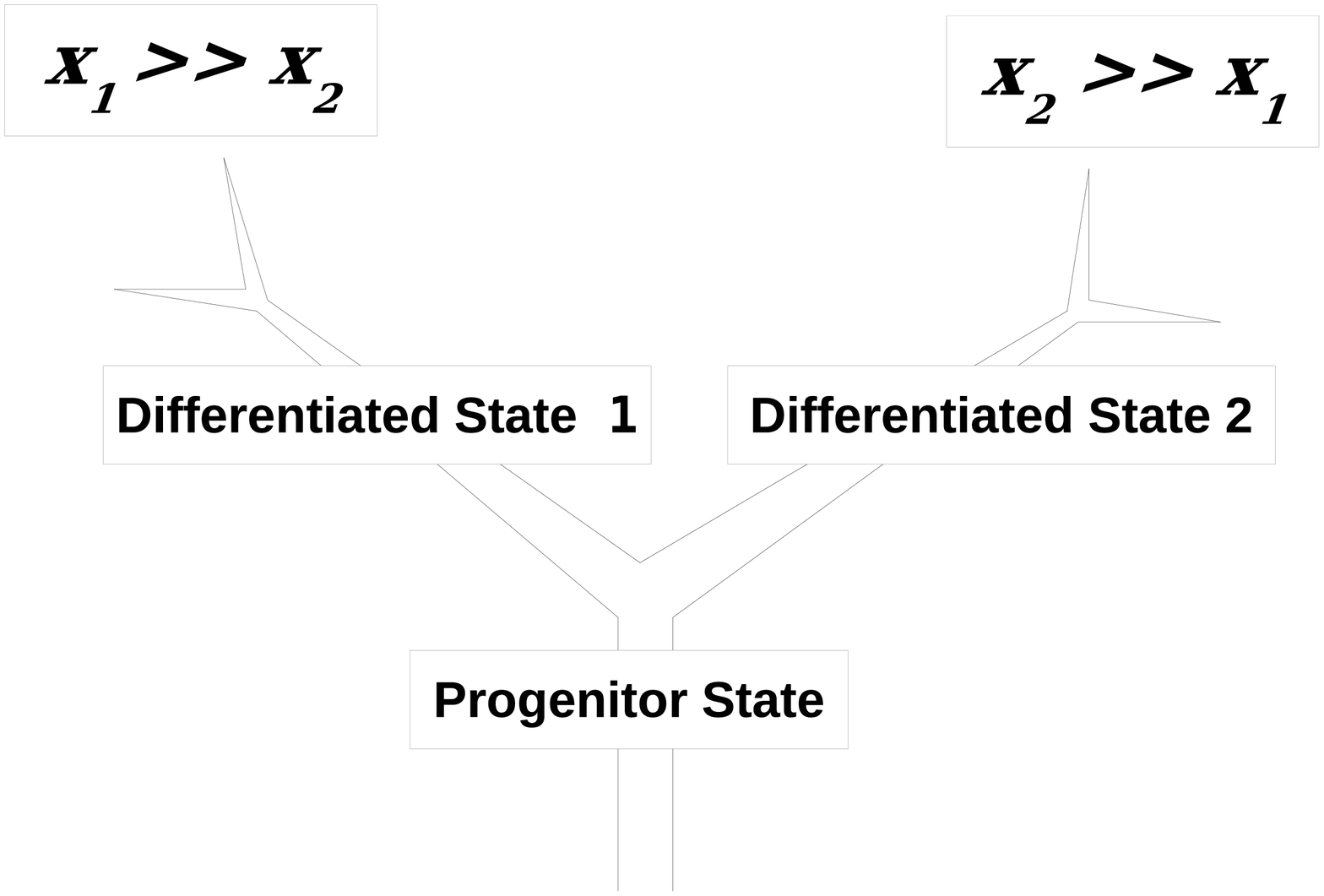}}

     \caption{\small  (i)The central motif of gene regulatory networks controlling cell differentiation. The protein products of two genes X{$_1$} and X{$_2$} 
     repress each other's synthesis, thus constituting a positive feedback loop. The proteins further autoactivate their own synthesis. (ii) Illustration of a progenitor state differentiating 
     into two states 1 and 2 distinguished by the distribution of the proteins {\it{x{$_1$}}}$\gg${\it{x{$_2$}}} and  {\it{x{$_2$}}}$\gg${\it{x{$_1$}}} \normalsize }
    \end{figure}
 signals, if any. In the long 
 time limit, the dynamics finally converge to an attractor in the state space. More than one attractor is possible with each attractor having its 
 own basin of attraction. The attractors define the steady states when all the temporal rates of change of the variables governing the cellular dynamics are zero. 
 The global dynamics of the gene regulatory network is best visualized in the landscape model. The landscape depicts a ``potential-energy-like'' function 
 (to be defined later) in the multi-dimensional state space and consists of valleys separated by hills. The valleys represent stable steady states, i.e, the 
 attractors of the dynamics whereas the hilltops represent the unstable steady states. In the context of cell differentiation, the stable steady states describe 
 the different cell types and the undifferentiated cell states are at the hilltops. Each valley defines the basin of attraction of the associated 
 steady state. A deeper valley represents a more stable steady state. The experimental evidence that the different cell types correspond to the 
 attractors of the gene regulatory network dynamics was provided by Huang and co-workers \cite{Huang}.
 
 Waddington's epigenetic landscape, the cell-fate tree and the landscape model have close parallels with one another in the depiction of the 
 multiple cell types arising due to cell differentiation. The gene regulatory network controlling the dynamics of cell differentiation is quite 
 complex but a key motif playing a central role in guiding the dynamics in a number of cellular systems has been identified \cite{Zhou,Furusawa}. The 
 motif consists of two genes X$_1$ and X$_2$ the protein products of which repress each other's expression (figure 1(i)). The proteins further autoactivate their 
 own production via individual positive feedback loops. The simple genetic circuit captures the binary decision process at a branch point of the cell-fate tree 
 (figure 1(ii)) or the choice between two valleys when an existing valley bifurcates into two valleys. The state in which the opposing genes have low and equal expression 
 levels x$_1$ $\approx$ x$_2$ corresponds to the undifferentiated state whereas the two stable steady states x$_1$ $\gg$ x$_2$ and 
 x$_1$ $\ll$ x$_2$, represent the differentiated cell states. In the undifferentiated cells, there is no dominance of one TF over the other. The 
 two differentiated lineages are distinguished by the reversal of expressions of the pair of antagonistic genes X$_1$ and X$_2$. A recent exhaustive 
 analysis of the human expression data involving 2602 transcription-regulating genes and 166 cell types provides substantial evidence that 
 the two-gene motif governs the differentiation of a progenitor cell into sister lineages in a large number of cases \cite{Heiniemei}.
 
 A number of mathematical models of cell differentiation have been proposed so far based on the two-gene motif and its variations 
 \cite{Huang1,Wang,Roeder,Chickermane,Bokes,Duff}. The most well-known of these is the minimal model developed by Huang and co-workers 
 \cite{Huang1,Wang} the dynamics of which describe both the autoactivation and the mutual inhibition of gene expression along with the protein 
 degradation. Some variants of the basic model have been proposed \cite{Roeder,Chickermane,Bokes,Duff} to take the experimentally observed 
 features into account. In this paper, we propose a model of cell differentiation which combines some of the realistic features of the existing 
 models. In our model, the two-gene motif \:\:( figure 1(i) ) forms the core of the gene regulatory network with the autoactivation and the 
 repression by the opposing TFs being non-cooperative in nature, i.e., mediated by monomers. Additionally, the model incorporates an indirect 
 repressive effect via the formation of a heterodimer of the two TFs. In section 2, we describe our model of cell differentiation and characterize 
 the different aspects of the dynamics by computing the bifurcation diagrams and analyzing the robustness of the dynamics to variations in the 
 parameter values. Recent experimental findings support the notion that gene expression noise in the form of fluctuations in the TF levels gives rise to 
 phenotypic heterogeneity in clonal cell populations \cite{Kaern,Raj,Balazsi}. The noise has a non-trivial role in cell-fate decisions like lineage 
 choice. This has been demonstrated in an experiment by Chang {\it{et al}} \cite{Chang} who observed the existence of considerable heterogeneity in 
 the levels of the stem-cell-surface marker protein Sca-1 in a clonal population of mouse hematopoietic ( blood cell forming ) progenitor cells. The 
 heterogeneity generates distinct differentiation biases in the cell population. Cells with low (high) Sca-1 levels predominantly differentiate 
 into the erythroid (myeloid) lineage. In sections 3 and 4 of the paper, we investigate the dynamics of our model in the presence of noise and analyze the 
 results obtained to provide a comprehensive physical understanding of the experimental results of Chang {\it{et al}} \cite{Chang}.
 
 Recently a large number of studies have been carried out on the early signatures of sudden regime shifts in systems as diverse as ecosystems, 
 financial markets, population biology and complex diseases \cite{Scheffer,Scheffer1}. Similar studies on the signatures of regime shifts in gene 
 expression dynamics have been carried out in Refs. \cite{Pal,Ghosh}. Regime shifts in general occur at bifurcation points or may be noise-induced. 
 The types of bifurcation considered in models of cell differentiation include the saddle node and pitchfork bifurcations 
 \cite{Wang,Roeder,Chickermane,Bokes, Duff,Strogatz}. A bifurcation brings about a regime shift from an undifferentiated cellular state to a differentiated cell lineage. Since the 
 dynamics associated with cell differentiation have a stochastic component, the levels of the TFs, playing key roles in the lineage choice, are described in terms 
 of probability distributions. The early signatures of sudden regime shifts, occurring at bifurcation points, include the critical slowing down 
 and an increase in the variance, lag-1 autocorrelation function and the skewness of the probability distribution as the bifurcation point is approached 
 \cite{Scheffer,Scheffer1}. These quantities are experimentally measurable and provide evidence of an impending regime shift. In 
 section 4 of the paper, we provide a quantitative characterization of cell differentiation as a bifurcation phenomenon in terms of specific early signatures. 
 Such signatures can further distinguish between alternative mechanisms of cell differentiation. In section 5, we discuss the significance of the major results 
 obtained in the paper.
 
 \section*{2. Model of cell differentiation}
 \label{Model}
 
 We propose a mathematical model of cell differentiation based on the two-gene motif of figure 1(i). The steady state protein concentrations for the two genes 
 X$_1$ and X$_2$ are represented by the variables {\it{x$_1$}} and {\it{x$_2$}}. The rate equations describing the dynamics of the model are:
 \begin{equation}
 \frac{ dx{_1}}{dt}\ = \frac{a_{1} x{_1}}{S+x{_1}}\ + 
   \frac{b_{1} S}{S+x{_2}}\ -k_{1} x{_1}- g x{_1} x{_2}
    \end{equation}
    
    \begin{equation}
    \frac{ dx{_2}}{dt}\ = \frac{a_{2} x{_2}}{S+x{_2}}\ + 
   \frac{b_{2} S}{S+x{_1}}\ -k_{2} x{_2}- g x{_1} x{_2}
   \end{equation}
   
   The first terms on the right hand sides of the two equations describe the autoactivation, the second terms represent the cross-repression of the 
   expression of the two genes, the third terms correspond to the individual protein degradation rates and the fourth terms represent the indirect 
   repression of gene expression via the heterodimer formation. The model differs from the model of Wang {{\it{et al}} \cite{Wang} in two respects: 
   ({\it{i}}) no cooperativity is involved in the autoactivation as well as the cross-repression (in the presence of cooperativity, the 
   regulatory molecules form bound complexes of n molecules ({\it{n}} $>$ 1) with n known as the Hill coefficient) and 
   ({\it{ii}}) the heterodimer formation by the proteins {\it{x$_{1}$}} and {\it{x$_{2}$}} is taken into account. These features have some experimental support 
   in the case of blood cell differentiation in a population of HSCs. Again, a two-gene motif plays a key role with the genes X$_1$ and X$_2$ synthesizing 
   the opposing TFs GATA1 and PU.1. The progenitor blood cells have two differentiated lineages, erythroid ({\it{{x$_1$}}} $\gg$ {\it{{x$_2$}}}) and 
    myeloid ({\it{{x$_1$}}} $\ll$ {\it{{x$_2$}}}) with  {\it{{x$_1$}}} and {\it{{x$_2$}}} denoting the TFs GATA1 and PU.1 respectively. There is,
   \begin{figure}[h]
   \centering
   \subfloat[\:\:( i )]{%
     \includegraphics[scale=0.25,width=0.19\textwidth]{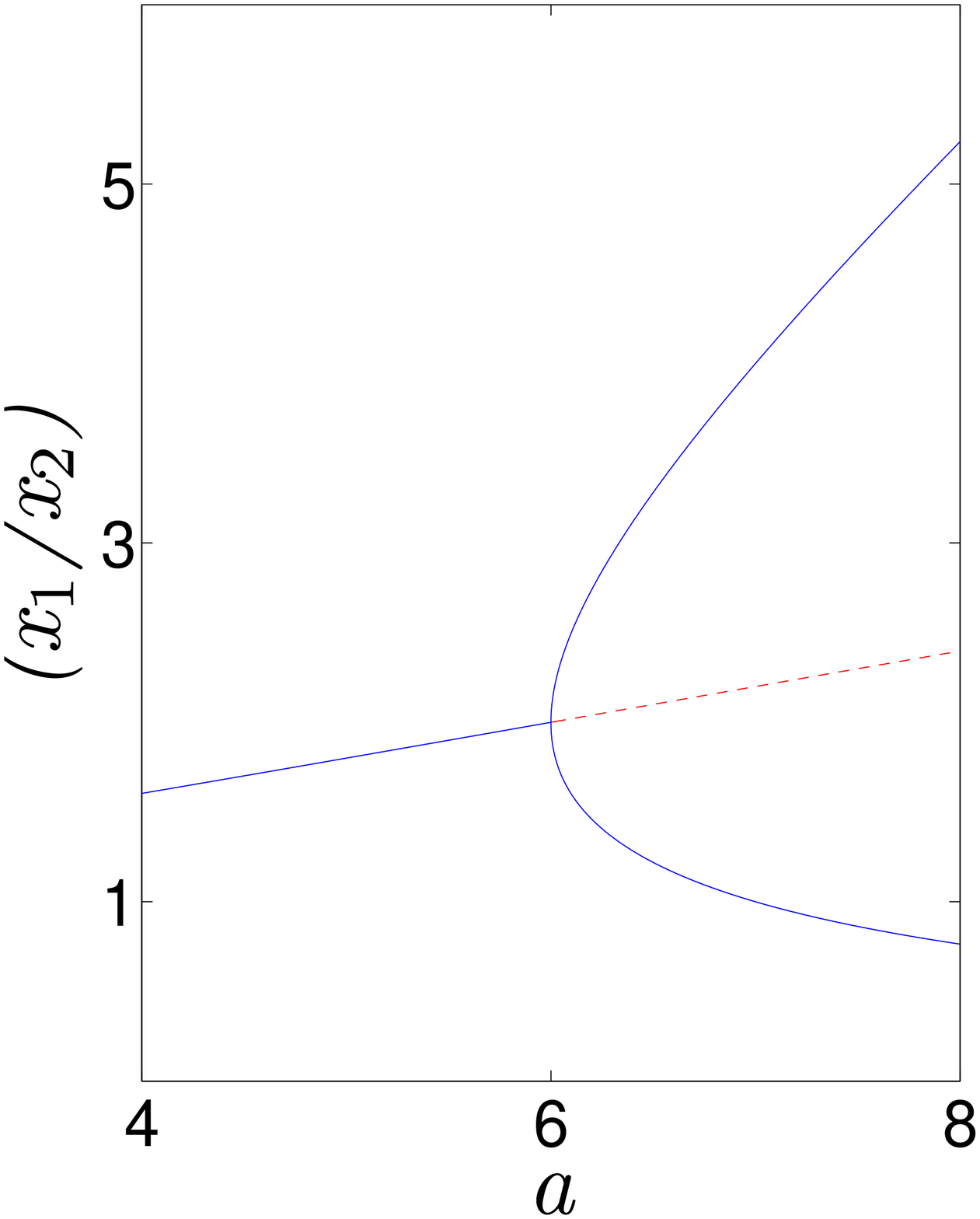}}
   \quad
   \subfloat[\:\:( ii )]{%
     \includegraphics[scale=0.25,width=0.19\textwidth]{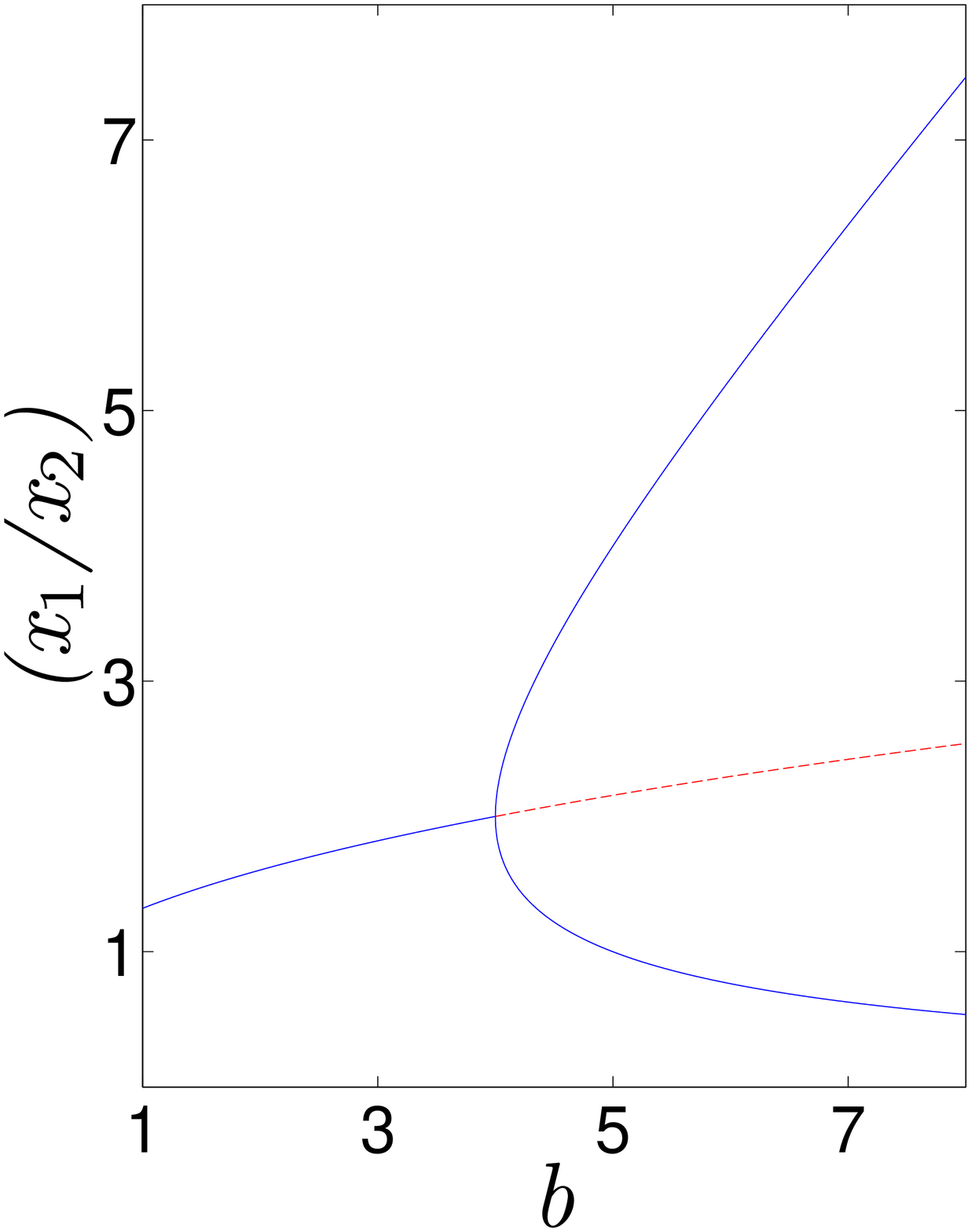}}
    
     \subfloat[\:\:( iii )]{%
     \includegraphics[scale=0.25,width=0.19\textwidth]{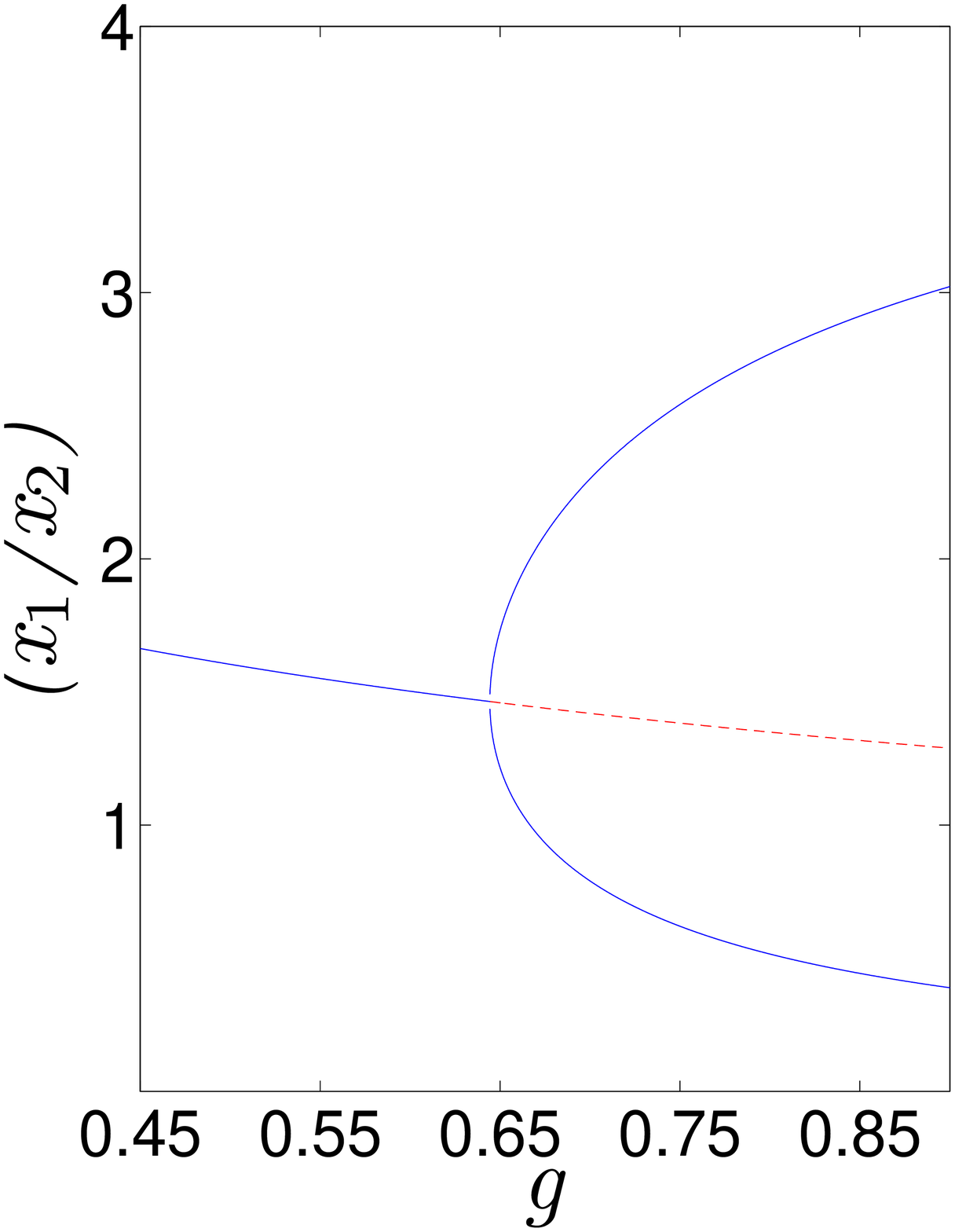}}
   \quad
   \subfloat[\:\:( iv )]{%
     \includegraphics[scale=0.25,width=0.19\textwidth]{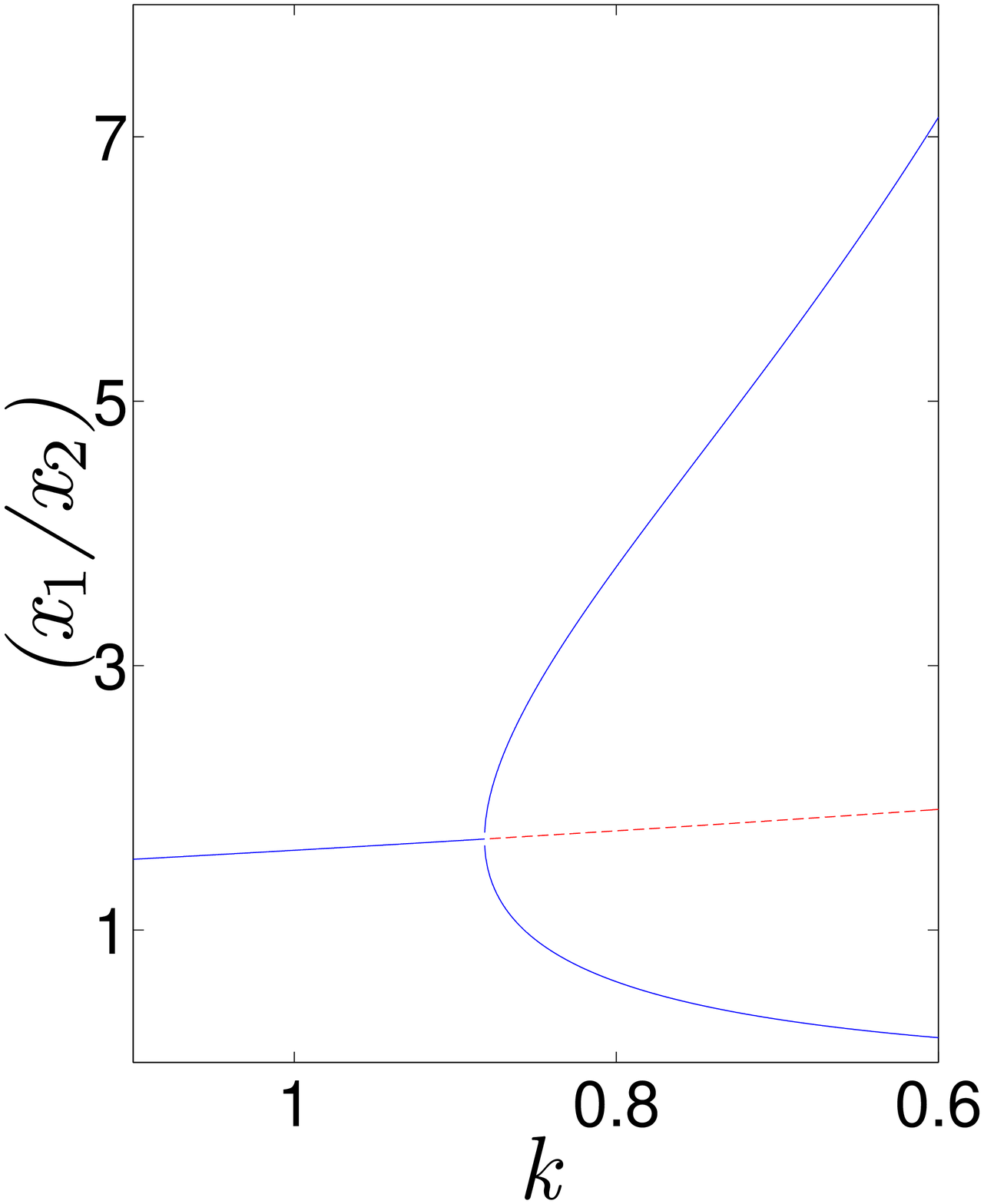}}
     \caption{\small Bifurcation diagrams depicting the steady state values {\it{x{$_1$}}} ({\it{x{$_2$}}}) versus different bifurcation parameters: 
     (i) {\it{a}}, (ii) {\it{b}}, (iii) {\it{k}} and (iv) {\it{g}}. The parameter values are listed in Table 1. The solid (dotted) lines represent stable (unstable) 
     steady states.\normalsize }
    \end{figure}
   so far, no conclusive evidence that the autoregulation of GATA1/PU.1 is mediated by dimers or higher order multimers. This has led to models in which 
   no cooperativity in the autoregulation is assumed, i.e., the autoregulation is mediated by monomers only \cite{Chickermane,Bokes}. In 
   our model, the mutual repression also lacks cooperativity. From experiments, the mutual antagonism between PU.1 and GATA1 appears to involve the 
   formation of the PU.1-GATA1 heterodimer \cite{Liew}. The heterodimer has an inhibitory function via two possible mechanisms 
   \cite{Roeder,Bokes,Duff}: ({\it{i}}) the heterodimers repress transcription initiation by binding the appropriate regions of the DNA 
   corresponding to the genes X$_1$ and X$_2$, ({\it{ii}}) the heterodimer does not bind the DNA. The first mechanism involves a competition 
   between the heterodimers and the autoactivating TFs for free binding sites. In the second case, the heterodimer formation restricts the availability 
   of the free TFs to autoactivate their own synthesis. This results in an indirect repression of the transcriptional activity. Our model is similar 
   to the model proposed by Bokes {\it{et al}} \cite{Bokes} with a few differences. In the latter
   model, the cross-repression of gene expression by the 
   opposing enzymes ( the second terms on the right hand sides of equations (3) and (4)) is ignored. The possibilities for heterodimer degradation 
   and the dissociation of a heterodimer into free monomers are further included in their model. The inclusion of heterodimer dissociation 
   in our model does not change the results obtained in a qualitative manner. In the model proposed by Duff {\it{et al}} \cite{Duff}, the heterodimer represses 
   the initiation of transcription in the cases of both the genes X$_1$ and X$_2$ and the autoactivation as well as the repression involve 
   cooperativity. In our model, the individual proteins, rather than the heterodimers, act as the direct 
   \begin{table}[t]
    \caption{ 
    Parameter values used in different figures}
    \centering
    \begin{tabular}{c c c c c c}
     \hline\hline
     Description & {\it{a}} & {\it{b}} & {\it{S}} & {\it{k}} & {\it{g}}\\
     \hline
     Figure 2 (i)\:\:\: & bifurcation & 2.0 & 2.0 & 1.0 & 0.5 \\[0.01ex]
     &\raisebox{0.01ex}{parameter}\\[0.01ex]
     &\raisebox{0.01ex}{(b.p.)}\\
     Figure 2 (ii) & 4.0 & b.p. & 2.0 & 1.0 & 0.5 \\[2ex]
     Figure 2 (iii) & 4.0 & 2.0 & 2.0 & b.p. & 0.5\\[2ex]
     Figure 2 (iv) & 4.0 & 2.0 & 2.0 & 1.0 & b.p.\\[2ex]
     Figure 3 \:\:\:\:\:\:\:& 6.0 & 6.0 & 2.0 & 1.0 & 0.5\\[0.2ex]
     \raisebox{0.1ex}{(reference set}\\[0.1ex]
     \raisebox{0.1ex}{of values)}\\[2ex]
     
     Figure 6 \:\:\:\:\:\:\:& {\it{a}} = 1.05 & 1.0 & 0.5 & 1.0 & 0\\[0.2ex]
     \raisebox{0.1ex}{(model of Wang}& \raisebox{0.1ex}{(region of }\\[0.2ex]
      \raisebox{0.1ex}{{\it{et.al}} [9]. The}\:\:\: &\raisebox{0.1ex}{tristability)}\\[0.001ex]
     \raisebox{0.1ex}{Hill coefficient}\\[0.2ex]
     \raisebox{0.1ex}{ {\it{n}} = 4}\\
     
    \hline              
    \end{tabular}

   \end{table}
    repressors of transcription. The latter repress the 
   transcriptional activity in an indirect manner. Rouault and Hakim \cite{Roualt} have proposed two simple gene circuits involving two genes to 
   demonstrate how identical cells choose different fates via cell-cell interactions. The two genes X$_1$ and X$_2$ now belong to two different cells. In Model 1, 
   the protein A autoactivates its own synthesis while the protein B is expressed constitutively ( no regulation ). The proteins A and B further 
   form hetrodimers ( hetrodimer dissociation and degradation are not considered ). In the case of Model 2, there is no autoactivation of protein synthesis and the protein A 
   represses the synthesis of the protein B. Both the proteins A and B are constitutively expressed and also form heterodimers. In specific parameter 
   regimes, both the models exhibit bistability with the stable steady states representing different cellular states.
   
   In the case of our model, we study the symmetric situation {\it{a$_1$}}={\it{a$_2$}}={\it{a}}, {\it{b$_1$}}={\it{b$_2$}}={\it{b}} 
   and {\it{k$_1$}}={\it{k$_2$}}={\it{k}}. Figure 2 shows the computed bifurcation diagrams for the steady state values of {\it{x$_1$}} ( {\it{x$_2$}} ) versus the bifurcation parameter 
   {\it{a}}, {\it{b}}, {\it{k}} and {\it{g}}. In all the cases, the solid lines represent the stable steady states and the dotted lines denote the unstable steady states. 
   Table 1 displays the parameter values for figures 2 (i) - (iv). In each case, a supercritical pitchfork bifurcation is obtained in which a monostable state loses stability 
   at the bifurcation point with the simultaneous appearance of two new stable steady states corresponding to the differentiated cell states. The bifurcation diagrams obtained by us are different 
   from those associated with the model studied by Wang {\it{et al}} \cite{Wang}. In the symmetric case of the latter model, the bifurcation diagram in terms of the parameter {\it{a}}  is characterised by a subcritical 
   pichfork bifurcation separating a region of tristability from a region of bistability. The central steady state of the three stable steady states describes the multipotent progenitor state whereas the 
   other stable steady states represent the differentiated states. The progenitor state loses stability in the region of bistability so 
   that only the differentiated states are present. A similar bifurcation diagram is obtained when the steady state protein level in plotted as a function of the 
   parameter {\it{k}} in the symmetric case. A supercritical pichfork bifurcation is, however, obtained in the case of the bifurcation parameter {\it{b}}.
   
   In the case of our model, there is no parameter region in which tristability, as in the case of the model proposed by Wang {\it{et al}} \cite{Wang}, is obtained. Tristability is an outcome of 
   cooperativity in the regulation of gene expression for which there is no substantial experimental evidence. A subcritical pitchfork bifurcation from tristability to bistability is achieved as the bifurcation parameter {\it{a}}, the strength of the autoregulation of gene expression, 
   is decreased. In the case of our model, a supercritical pitchfork bifurcation occurs as
   the bifurcation parameter {\it{a}} is increased. If cell differentiation is an outcome of 
   \begin{figure}
   \centering
  \includegraphics[scale=0.35]{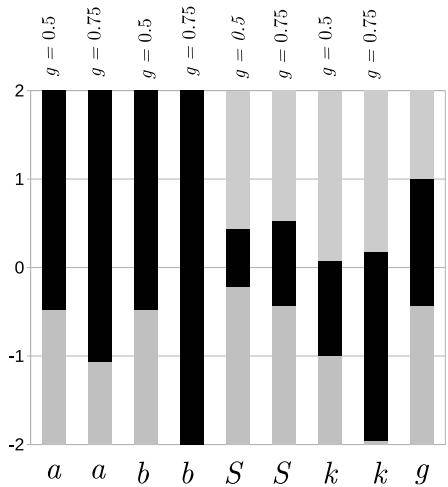} 
  \caption {\small Analysis of robustness of parameters in equations (1) and (2). The value of each parameter has a 100-fold variation greater and lesser than the reference value 
  shown in Table 1. The x-axis shows the parameter varied and the y-axis depicts the log-fold variation. The region of bistability is shaded dark.
   \normalsize }
   \end{figure}
   a bifurcation, 
   then in the first (second) case the cell differentiation occurs as a decreasing (increasing) function of the parameter {\it{a}}. In the absence of cooperativity in the case of the model of Ref. \cite{Wang}, 
   there is no pichfork bifurcation, either subcritical or supercritical. The fundamental requirement for the generation of bistability in a dynamical system is the presence of positive feedback combined with an ultrasensitive (sigmoidal) 
   response \cite{Chen}. The most common origin of ultrasensitivity lies in cooperativity, e.g., in the regulation of gene expression. A less studied source of ultrasensitivity involves protein 
   sequestration \cite{Buchler,Buchler1} in which the activity of a protein A is compromised through sequestration via the formation of an inactive complex with another protein B. The concentration of the free active A exhibits an 
   ultrasensitive response with the threshold set by the condition A{$_T$} =  B{$_T$}, where A{$_T$} and B{$_T$} are the total concentrations of the protein A and its inhibitor B respectively. For A{$_T$} $<$ B{$_T$}, the concentration of free active A is 
   negligible as most of the molecules from complexes with the B molecules. For A{$_T$} $ >$ B{$_T$} and close to the threshold, the variation of the concentration of A versus A{$_T$} becomes ultrasensitive with a small change in A{$_T$} giving rise to a large change 
   in the concentration of free A. 
   \begin{figure}
   \subfloat[( i )]{%
    \includegraphics[scale=0.2,width=0.21\textwidth]{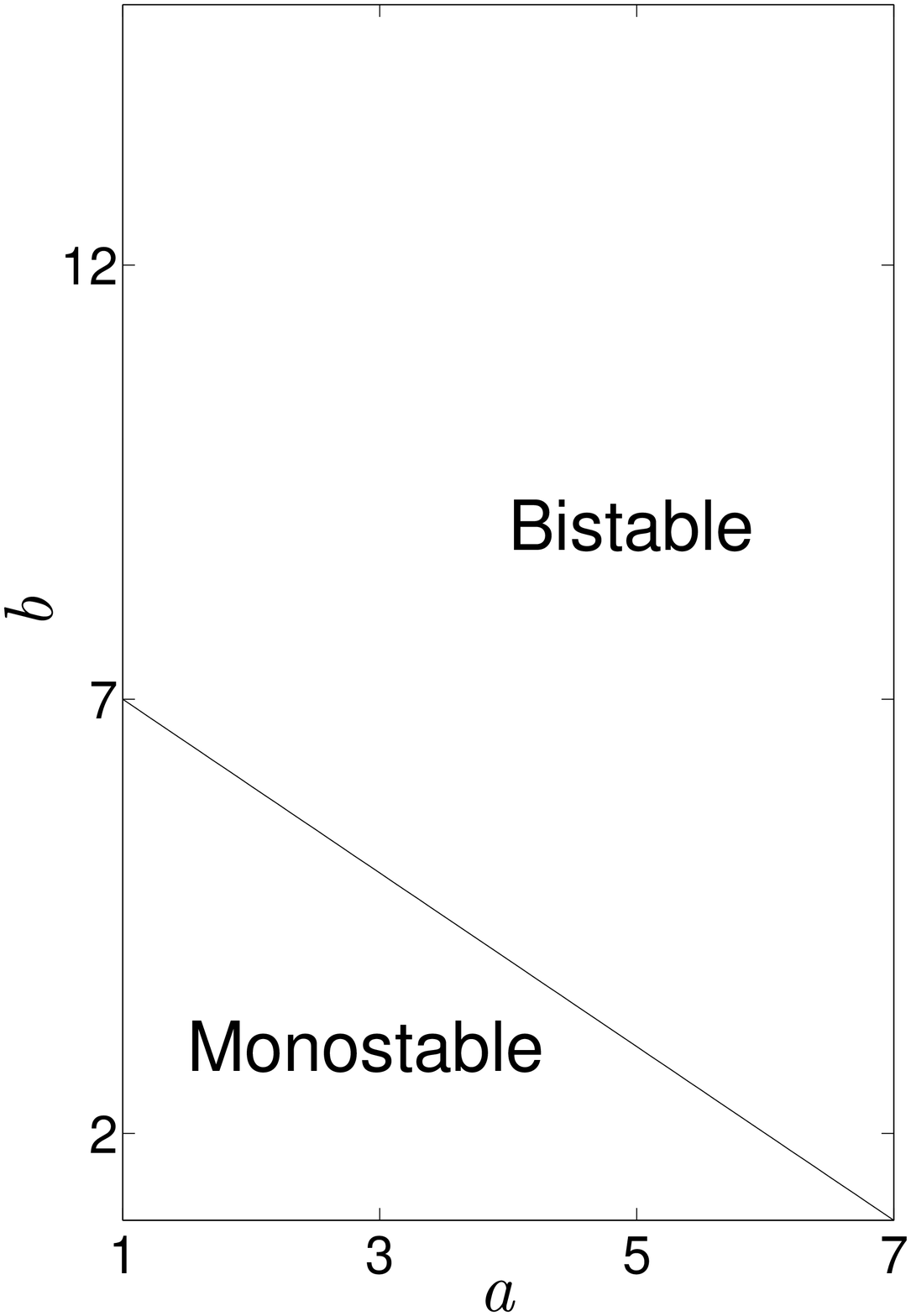}}
   \quad
   \subfloat[( ii )]{%
     \includegraphics[scale=0.2,width=0.21\textwidth]{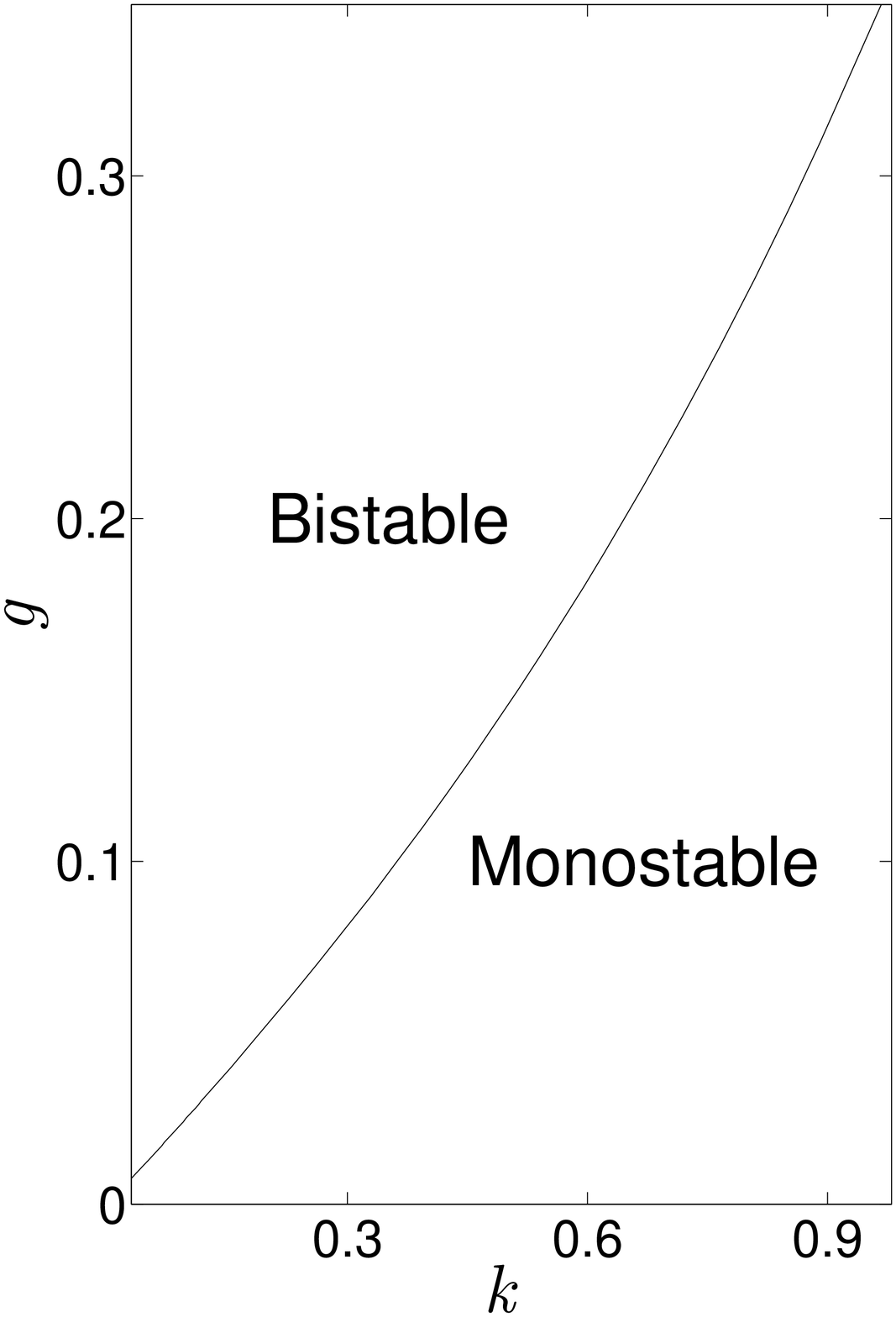}}
  \caption {\small Phase diagrams of our model for two-parameter scans: (i) {\it{a}} versus {\it{b}}, (ii) {\it{g}} versus {\it{k}}. The other parameter values are  from the reference set in Table 1.
   \normalsize }
   \end{figure}
   There are now several known examples of natural and synthetic biological systems which support bistability \cite{Chen,Buchler,Buchler1}. In most of the cases, the ultrasensitivity is an outcome of cooperativity though 
   sequestration-based ultrasensitivity has also been demonstrated in some of the systems. Chen and Arkin \cite{Chen} were the first to demonstrate experimentally that positive feedback along with sequestration were sufficient to construct a bistable switch. Reverting back 
   to the models of cell differentiation, the model proposed by Wang et al. \cite{Wang} relies on cooperativity to generate ultrasensitivity whereas the model studied by us is characterized by sequestration-based ultrasensitivity. In figure 3, we display the results of a parameter 
   sensitivity analysis with respect to the parameters {\it {a}}, {\it{b}}, S, {\it{k}} and {\it{g}}. The set of parameter values displayed in Table 1 serves as the reference set. In the case of the parameter {\it{g}}, another reference value {\it{g}}=0.75 is also used to determine the effect of increased {\it{g}} on the 
   extent of the region of bistability. The value of each of the parameters was varied separately in steps up to values that are 100 fold greater and lower than the reference value listed in Table 1. The figure shows the log-fold change in parameter value within the range -2 to +2 for a specific parameter (marked on the x axis ) and identifies the region of 
   bistability in the parameter range. We find that bistability occurs over an extended parameter regime. Also, the region of bistability increases in extent as the parameter {\it{g}} is increased in value. In figure 4, we show the phase diagrams for two-parameter scans, {\it{a}} 
   versus {\it{b}} (figure 4(i)) and {\it{g}} versus {\it{k}} (figure 4(ii)) in which the regions of monostability and bistability are identified. The other parameter values are from the reference set given in Table 1. Again, we find that our model exhibits bistability in an extended region of parameter space.   
   \section*{3. Stochastic dynamics}
   
   In the case of fully deterministic dynamics, cell differentiation involves bifurcation, i.e., an explicit external signal-driven parameter change is required. In this scenario, the stem cell population before differentiation is expected to be homogeneous, i.e., the individual cells in 
   the population are expected to have similar gene expression profiles. Recent experimental observations, however, indicate that heterogeneity is a characteristic feature of both the embryonic and the adult stem cells \cite{Graft}. 
   Interestingly, the heterogeneity turns out to be non-genetic in nature as established in experiments on clonal cell populations. The heterogeneity arises due to the randomness inherent in key biological processes, e.g., gene expression. Stochastic fluctuations ( noise ) in the TF levels give rise to a 
   broad distribution of protein levels in a population of cells. There is now experimental evidence that both deterministic dynamics and noise contribute 
   to cell-fate decisions \cite{MacArthur,Enver,Wang}. The experiment by Chang {\it{et al}} \cite{Chang} provides the evidence for an important functional role of non-genetic heterogeneity in cell differentiation. In the experiment, a clonal population of mouse hematopoetic progenitor cells is found to exhibit a nearly 
   1000-fold range in the levels of a stem-cell-surface marker protein Sca-1, i.e., the histogram of protein levels in the cell population is considerably broad. Using the fluorescence-activated cell sorting (FACS) technique, the `outlier' cells with very low and high levels of Sca-1 were sorted separately and cultured 
   in standard growth medium. The narrow Sca-1 histograms, obtained immediately after cell sorting were found to revert back to the parental distribution of Sca-1, the distribution before the sorting, over a period of two weeks. The slow kinetics constitute an important aspect of the progenitor cell dynamics. Another significant experimental observation was that the Sca-1 outliers have distinct transcriptomes. The cells with low (high) Sca-1 levels were found to 
   have high (low) GATA1 and low (high) PU.1 levels. The  outliers cells hence have distinct preferences for the choice of cell fate with the low Sca-1 subpopulation predominantly differentiating into the erythroid lineage because of the presence of high GATA1 levels in the subpopulation. Similarly, the high Sca-1 subpopulation 
   has a greater propensity towards choosing the myeloid lineage. The heterogeneity thus helps in the selection of a specific lineage through the `multilineage priming' of the progenitor population. The differently biased cells in the population attain their distinct cell fates in the presence of appropriate external signals.
   
   The experimental results of Chang {\it{et al}} \cite{Chang} are of significant importance in acquiring a physical understanding of how cell differentiation occurs in a large class of cell
   systems. The physical origins of the experimental observations are, however, still under debate \cite{Brock,Huang2}. In this section, we investigate the stochastic dynamics of our model and demonstrate that our model provides an appropriate physical basis for interpreting some of the experimental observations. 
   We study the stochastic dynamics of our model using the formalism based on the Langevin equations \cite{Fox,Van}:
     \begin{equation}
     dx{_1} = f(x{_1},x{_2})\;dt +\Gamma\;dW{_1}
    \end{equation}
    
    \begin{equation}
     dx{_2} = g(x{_1},x{_2})\;dt +\Gamma\;dW{_2}
   \end{equation}
   The functions f({\it{x{$_1$}}}, {\it{x{$_2$}}}) and g({\it{x{$_1$}}}, {\it{x{$_2$}}}) are given by the right hand side expressions in the equations (1) and (2) respectively, i.e., they are the same functions that describe the deterministic dynamics. W{$_1$} and W{$_2$} define the Wiener processes, which are independent white noise processes. The noise terms represent additive noise 
   with strength $\Gamma$, assumed to be the same in each case. Following standard procedure \cite{Fox,Foster}, the stochastic time evolution is computed using the update formulae with a time step $\Delta${\it{t}},
   \begin{equation}
     x{_1}(t+\Delta t) = x{_1}(t) + f(x{_1},x{_2})\Delta t + \Gamma \xi{_1}\sqrt{\Delta {\it t}} 
    \end{equation}
   
    \begin{equation}
     x{_2}(t+\Delta t) = x{_2}(t) + g(x{_1},x{_2})\Delta t + \Gamma  \xi{_2}\sqrt{\Delta {\it t}} 
   \end{equation}
   where $\xi${$_1$} and $\xi${$_2$} are Gaussian random variables with zero mean and unit variance.
   
   Figure 5 exhibits the results of the simulation of the Langevin equations (3) and (4) for $\Delta${\it{t}} = 0.01 and {\it{a}} = 5.98 (6.4) in the case of figure 5(i) ( figures 5(ii) - 5(iv) ). The other parameter values are the same as in the case of figure 2(i) (Table 1). 
   The steady state probability distributions in figure 5 are obtained after 50,000 time steps and over an ensemble of 10,000 cells. The noise strength is fixed to be $\Gamma$ = 0.07. The parameter value 
   {\it{a}} = 5.98 corresponds to the monostable, i.e, the undifferentiated cell state in figure 2(i). The steady state probability distribution P{$_{st}$}({\it{x{$_1$}}}, {\it{x{$_2$}}}) in this case is shown in figure 5(i). The `potential-energy-like' function mentioned in the Introduction in the context of the landscape picture is given by
   U ({\it{x{$_1$}}}, {\it{x{$_2$}}})\:$\sim$\: $-$ ln P{$_{st}$}({\it{x{$_1$}}}, {\it{x{$_2$}}}) and has a rugged character.
   
   The considerably broad probability distribution is consistent with experimental observations. We carry out a numerical experiment to mimic the experimental sorting of the clonal population into the low, medium and high Sca-1 subpopulations and demonstrate multilineage priming. Figure 5(ii) for the steady state probability 
   distribution has been obtained for {\it{a}} = 6.4, i.e., beyond the supercritical
   \begin{figure}[h]
   \centering
   \subfloat[( i )]{%
     \includegraphics[scale=0.2,width=0.21\textwidth]{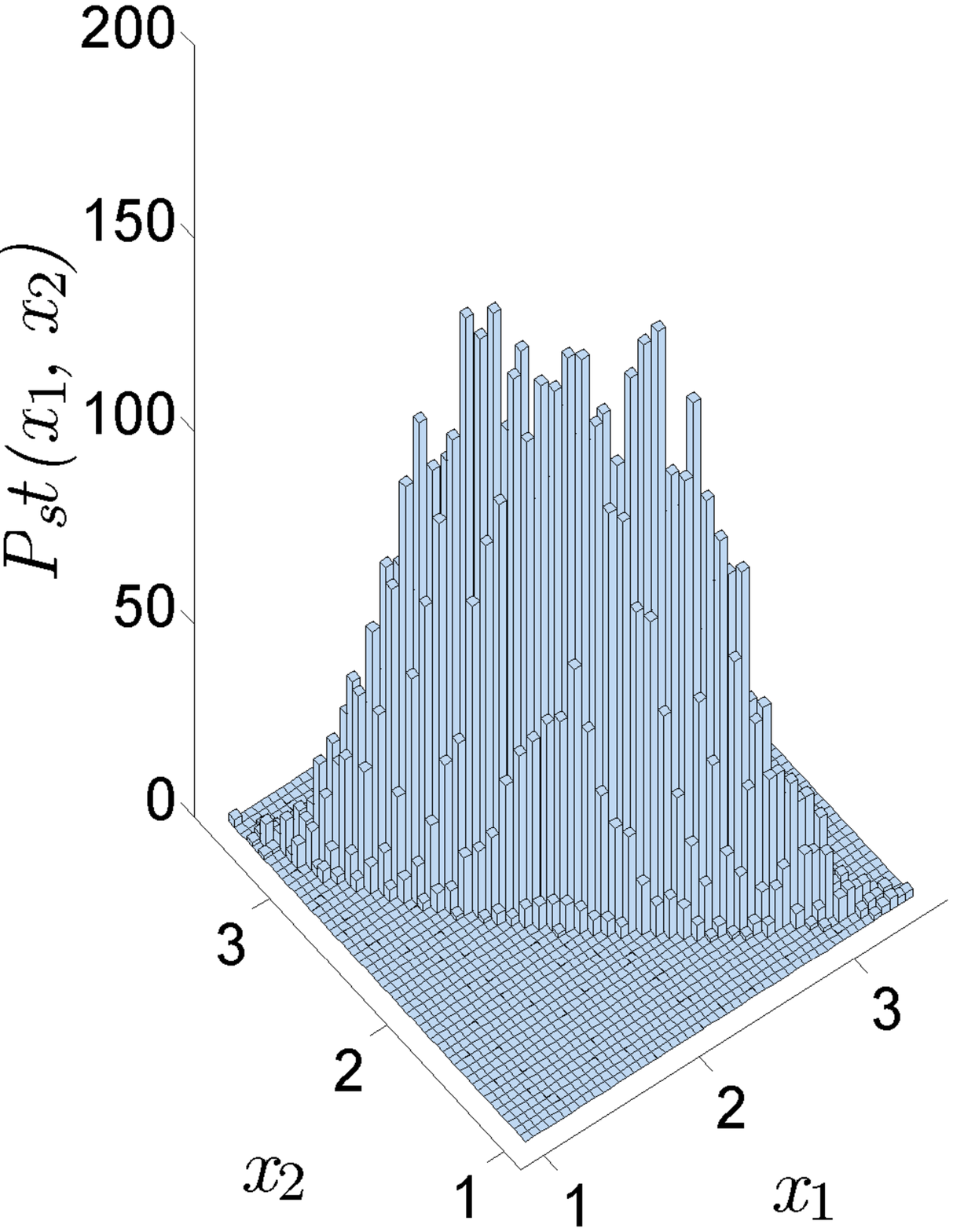}}
   \quad
   \subfloat[( ii )]{%
     \includegraphics[scale=0.2,width=0.21\textwidth]{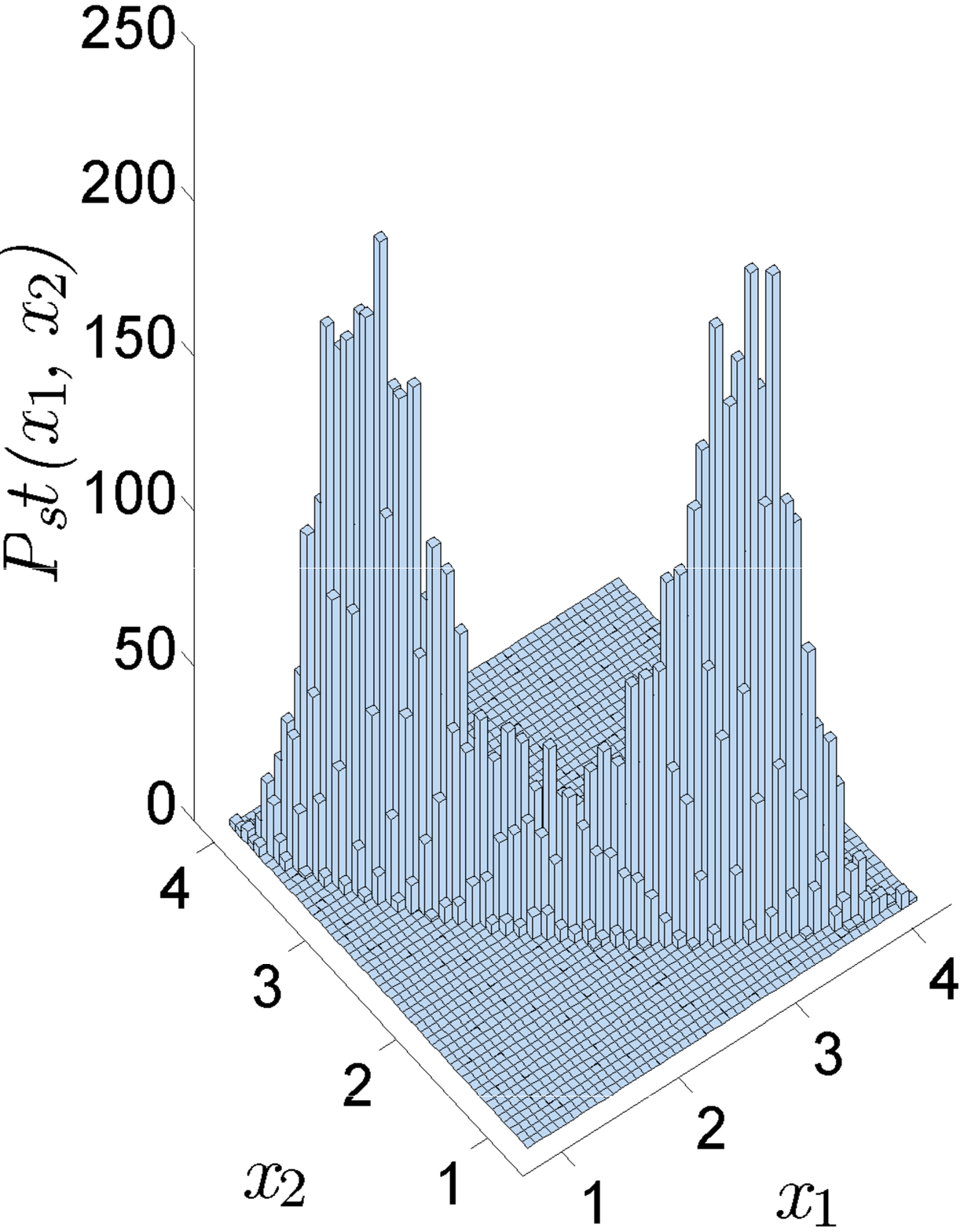}}
   
   \subfloat[( iii )]{%
     \includegraphics[scale=0.2,width=0.21\textwidth]{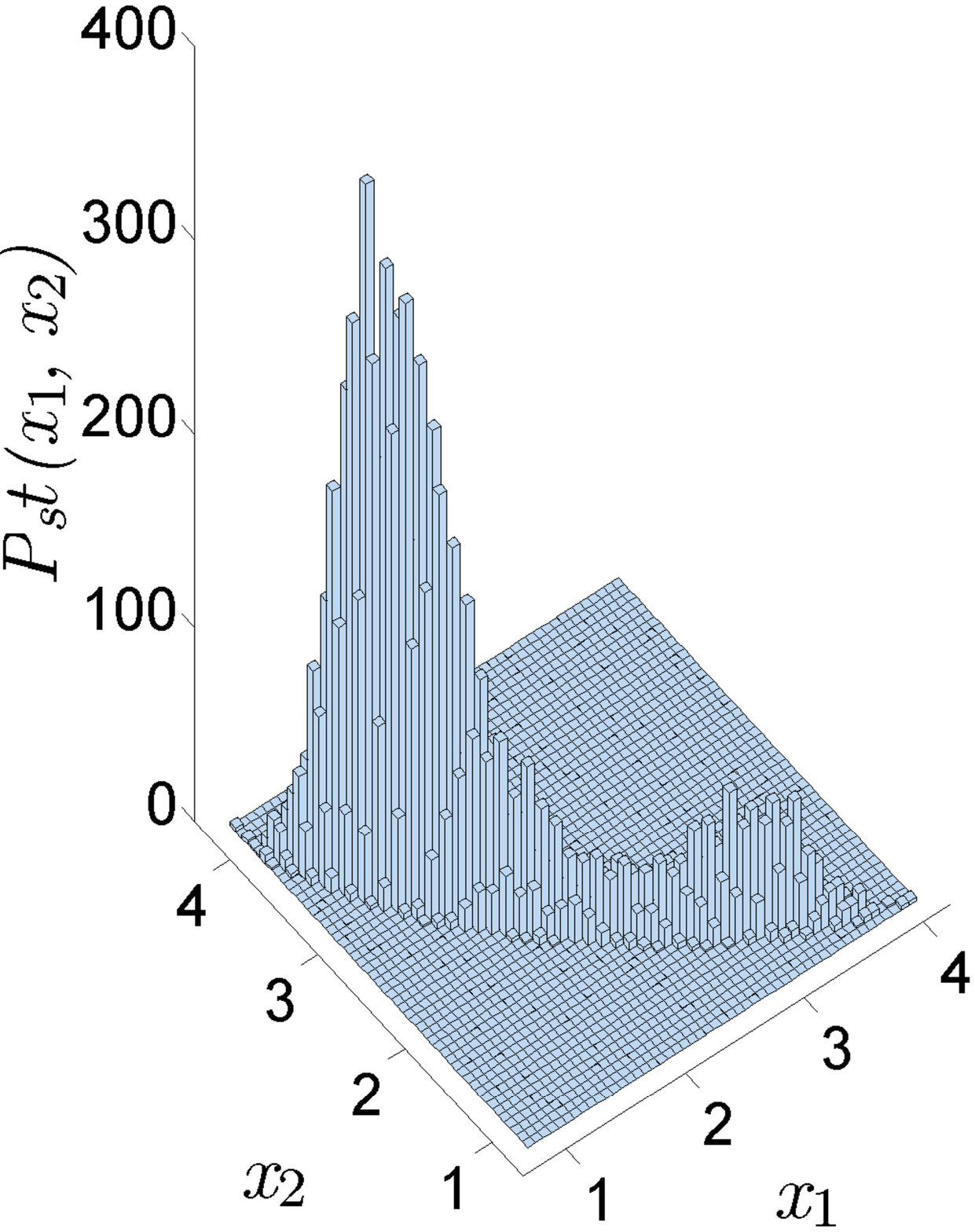}}
   \quad
   \subfloat[( iv )]{%
     \includegraphics[scale=0.2,width=0.21\textwidth]{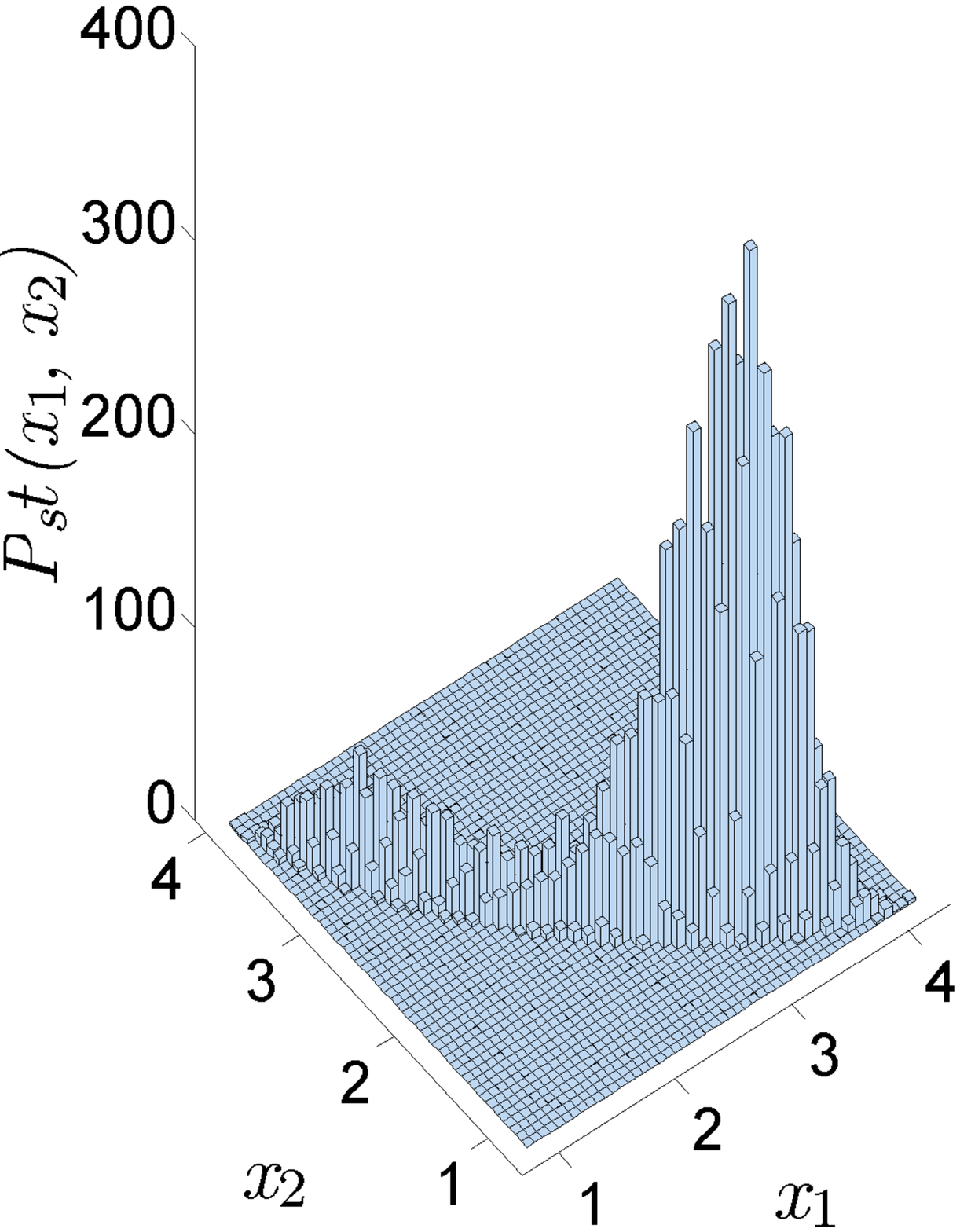}}
     \caption{\small Steady state probability distributions  P$_{st}$({\it{x{$_1$}}}, {\it{ x{$_2$}}}) obtained by simulating the Langevin equations (3) and (4) containing only 
     additive noise. ( i ) The parental histogram of the protein levels {\it{x{$_1$}}} and {\it{x{$_2$}}} in the case of undifferentiated cells. ( ii )-( iv ) Steady state 
     probability distributions for differentiated cells. The forms of the distributions are dictated by the locations of the initial states in the parental distributions. 
     The parameter {\it{a}} serves as the bifurcation parameter. \normalsize }
      \label{fig4}
     \end{figure}
   pichfork bifurcation point (figure 2(i)). The initial state corresponds to {\it{x{$_1$}}} = 2.0 and {\it{x{$_2$}}} = 2.0, i.e., belongs to the central region of the parental histogram. The two distinct peaks in P{$_{st}$} ({\it{x{$_1$}}}, {\it{x{$_2$}}}) 
   correspond to the subpopulations of differentiated cells. In the deterministic case, the choice of a stable steady state in the region of bistability is dictated by the initial state, specifically, the basin of attraction in the state space to which it belongs. In the presence of noise, the certainty of reaching a particular attractor state is lost and one can only speak of the propensity to attain the state. The initial 
   states corresponding to the extreme outlier cells in the parental distribution have distinct propensities towards specific lineages. This is illustrated in figures 5(iii) and 5(iv) corresponding to the initial states {\it{x{$_1$}}} = 1.4, {\it{x{$_2$}}} = 2.5 and {\it{x{$_1$}}} = 2.5, {\it{x{$_2$}}} = 1.4 respectively. Before the bifurcation point is crossed, there is only one basin of attraction so that a single peaked 
    steady state distribution is obtained whatever be the initial state. Beyond the bifurcation point, there are two basins of attraction and this is reflected in the two-peaked nature of the steady state probability distribution. For low levels of noise, the deterministic picture continues to be valid though instead of a single level, a distribution of levels is obtained in the steady state. 
    The states corresponding to the peaks of the distribution are representative of the stable steady states of the deterministic dynamics. The state {\it{x{$_1$}}} = 2.5, {\it{x{$_2$}}} = 1.4 belongs to the basin of attraction of the stable steady state with {\it{x{$_1$}}}$ > ${\it{x{$_2$}}}. This is reflected in the steady state probability distribution shown in figure 5(iv) with the cells being predominantly in the differentiated subpopulation of 
    lineage 1 ( {\it{x{$_1$}}} $ >$ {\it{x{$_2$}}} ). This is the erythroid lineage with the level of GATA1 ({\it{x{$_1$}}}) higher than that of PU.1 ({\it{x{$_2$}}}). Figure 4(iii) shows the steady state probability distribution when the lineage 2 is chosen preferentially by a cell ( {\it{x{$_2$}}} $>$ {\it{x{$_1$}}} ).
    
    We next repeat the numerical simulation as described above but instead of additive noise we consider multiplicative noise, the noise being associated 
    with the rate constant {\it{g}}, i.e., {\it{g}} $\rightarrow$ {\it{g}} + $\varepsilon$ where $\varepsilon$ denotes the random fluctuations in the rate constant. 
    The stochastic time evolution is computed using the update formulae similar to
   \begin{figure}[t]
   \centering
   \subfloat[( i )]{%
     \includegraphics[scale=0.2,width=0.21\textwidth]{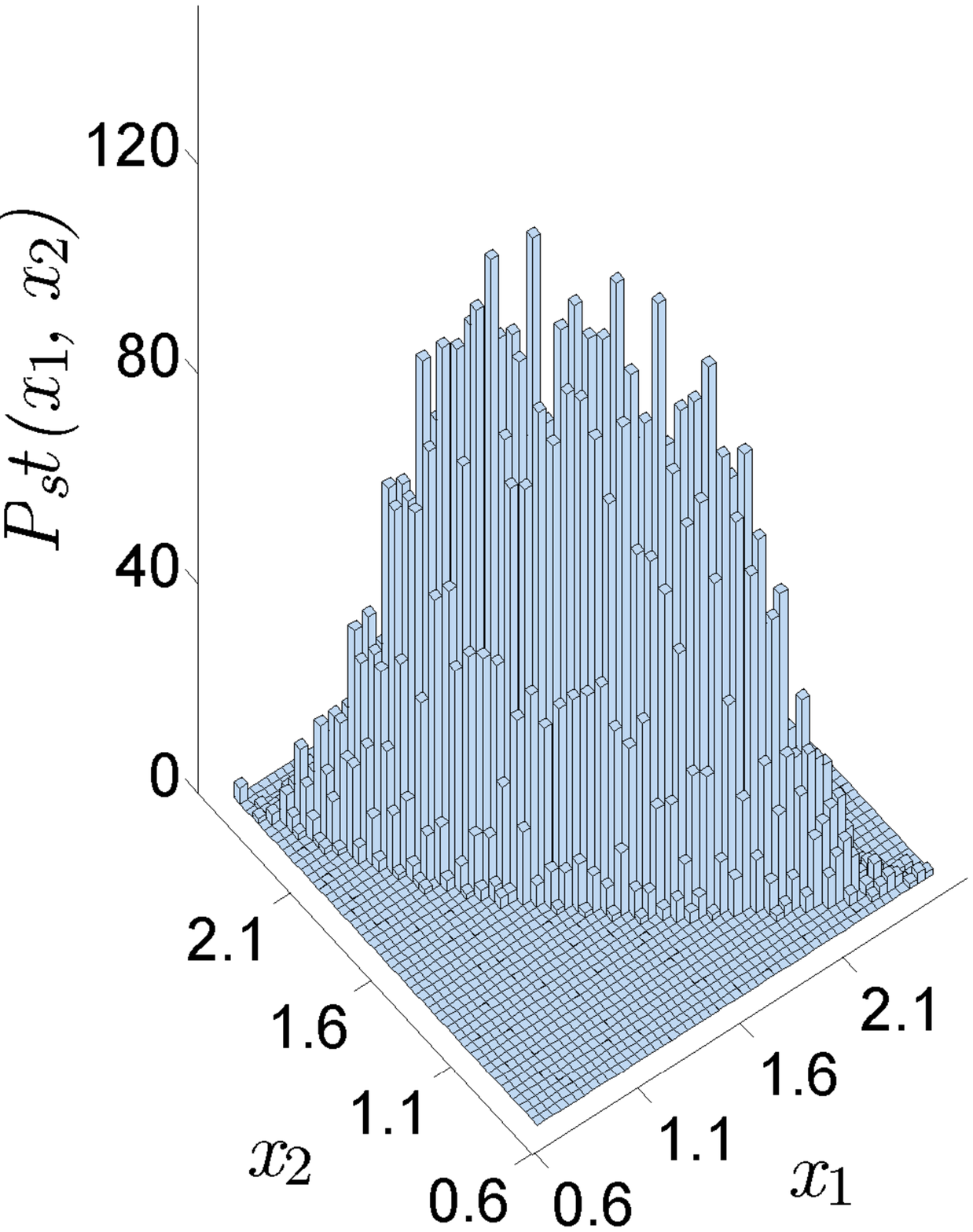}}
   \quad
   \subfloat[( ii )]{%
     \includegraphics[scale=0.2,width=0.21\textwidth]{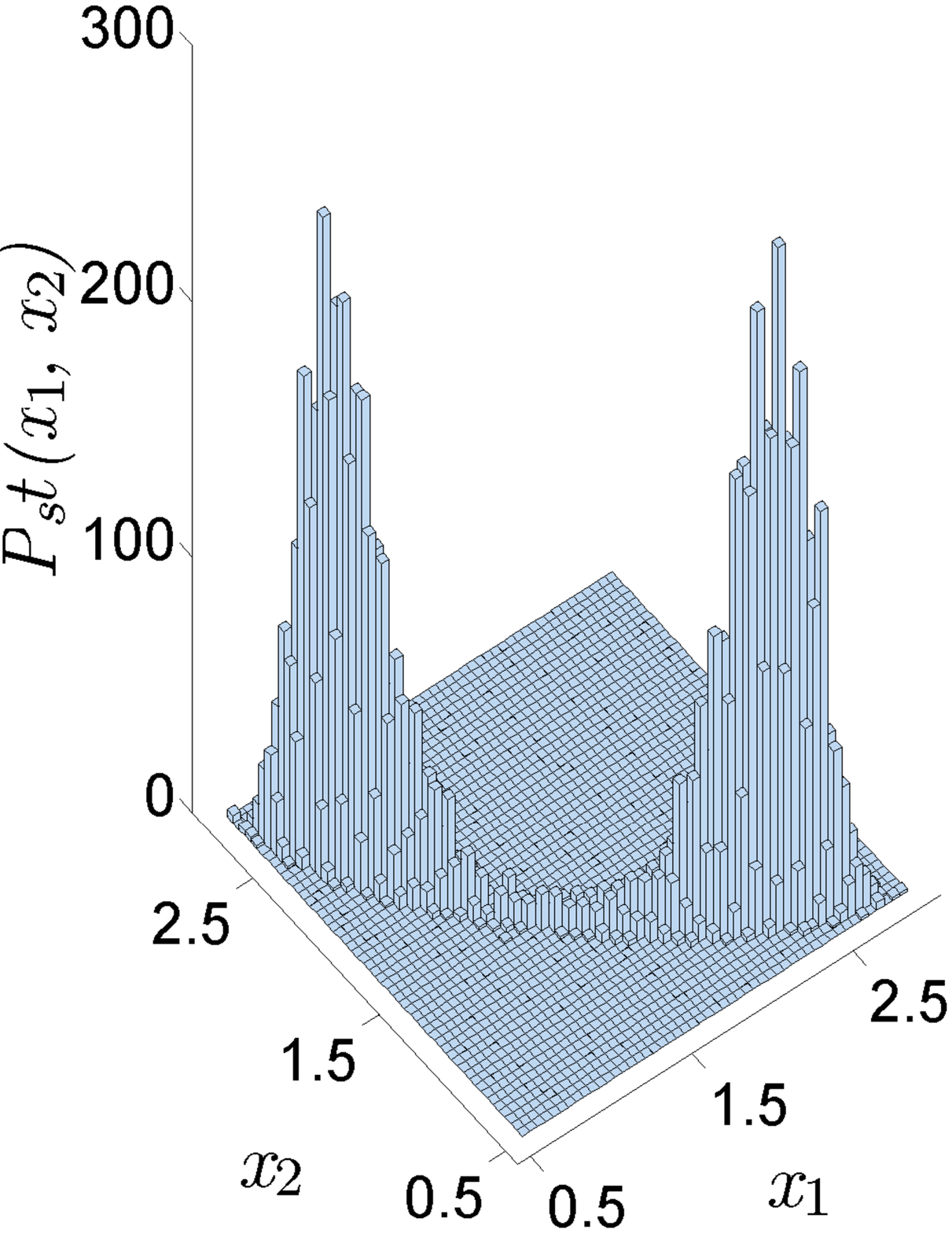}}
   
   \subfloat[( iii )]{%
     \includegraphics[scale=0.2,width=0.21\textwidth]{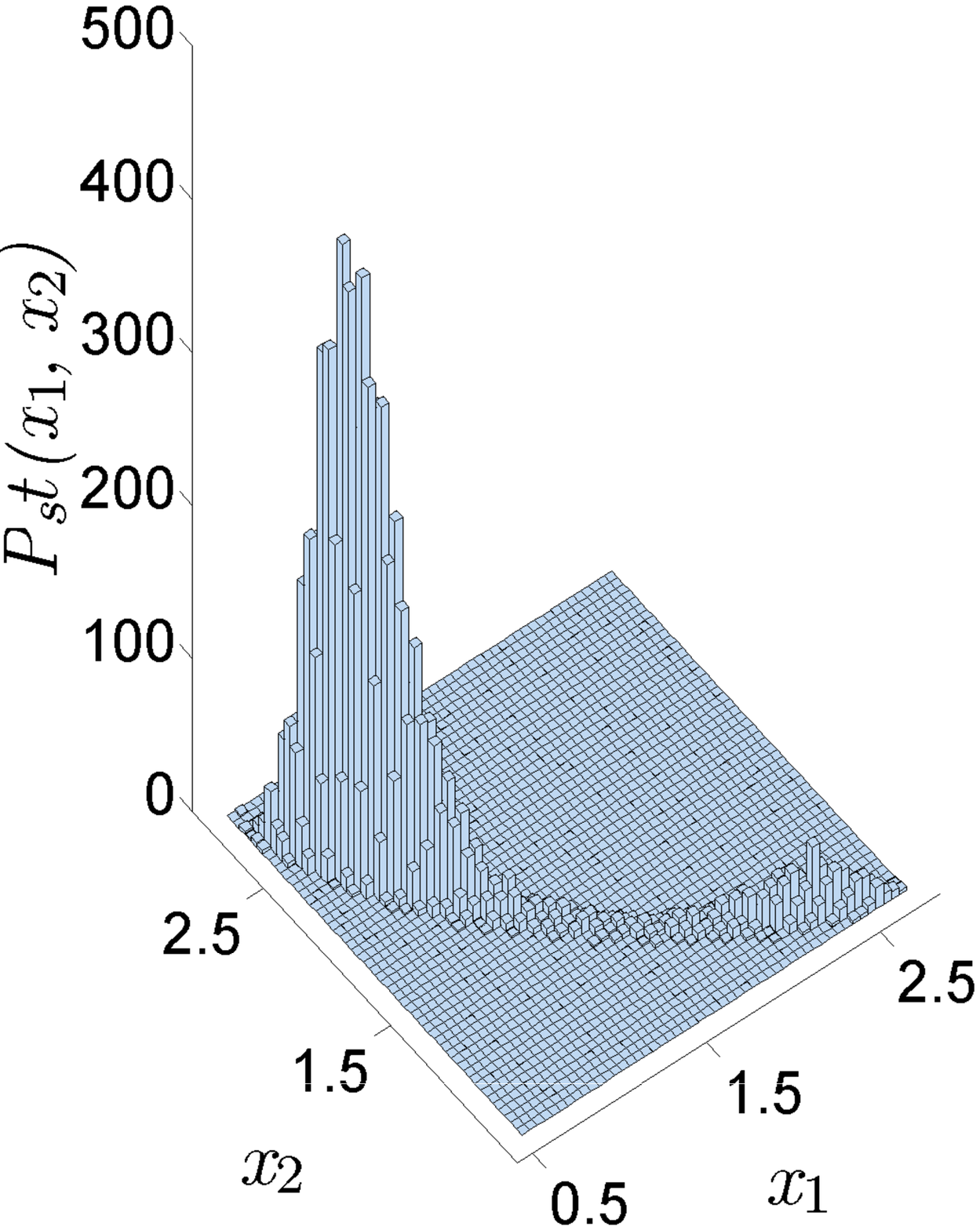}}
   \quad
   \subfloat[( iv )]{%
     \includegraphics[scale=0.2,width=0.21\textwidth]{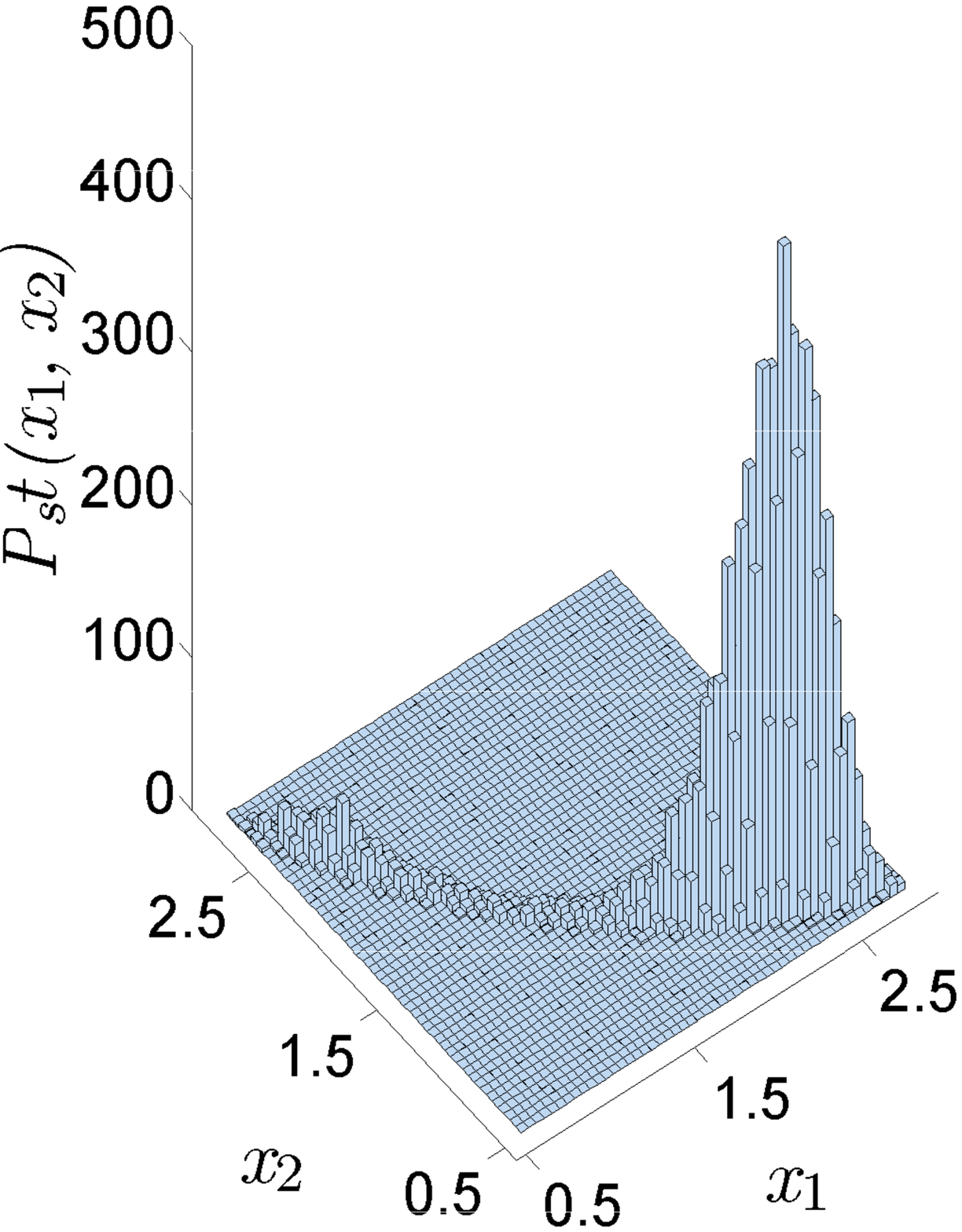}}
     \caption{\small Steady state probability distributions  P$_{st}$({\it{ x{$_1$}}}, {\it{ x{$_2$}}}) obtained by simulating the Langevin equations in the presence of 
     multiplicative noise in the parameter {\it{g}}. (i) The parental histogram of the protein levels in the case of undifferentiated cells. (ii)-(iv) 
     Steady state probability distributions for differentiated cells. The forms of the distributions are dictated by the locations of the initial states 
     in the parental distributions. The parameter {\it{g}} serves as the bifurcation parameter.\normalsize }
     \end{figure}
    those in equations (3) and (4) except that in the noise terms, $\Gamma$ is 
    replaced by \:\: $-$ $\varepsilon$ {\it{x}} ({\it{t}} ) {\it{y}} ({\it{t}} ). The steady state probability distributions are displayed in figure 6 with the rate constant {\it{g}} serving as the 
    bifurcation parameter. The parameter \:\:{\it{g}} = 0.642 (0.70) in the case of figure 6(i) (figures 6(i)-6(iv)). The other parameter values are the same as in the case 
    of figure 2(iv) ( Table 1 ). The noise parameter $\varepsilon$ has the value $\varepsilon$ = 0.035. Figure 6(i) describes the pre-bifurcation steady 
    state distribution whereas figures 6 (ii)-(iv) correspond to the post-bifurcation scenario. The initial state is {\it{x{$_1$}}} = 1.5, {\it{x{$_2$}}} = 1.5 in the 
    cases of figures 6 (i) and (ii). In the other two cases, the initial states are {\it{x{$_1$}}} = 0.8, {\it{x{$_2$}}} = 2.0 ( figure 6(iii) ) and {\it{x{$_1$}}} = 2.0, {\it{x{$_2$}}} = 0.8 
    ( figure 6(iv)). Both figures (5) and (6) illustrate conclusively the multilineage priming of the undifferentiated (parental) cell population which is a key experimental observation \cite{Chang}. The priming 
    of the parental cell population occurs irrespective of the nature of the noise, additive or multiplicative.
     
    Our model of cell differentiation provides a physical explanation of the origin of multilineage priming. The relaxation of the sorted `outliers' 
    to the parental distribution of Sca-1 levels in the experiment, supports the idea that the multipotent (undifferentiated) state is an attractor of the 
    dynamics. In the pre-bifurcation case this is the only attractor of the dynamics irrespective of the location of the initial state in the state space. 
    Beyond the bifurcation point, there are two stable attractor states the choice between which is determined by the location of the initial state in either of the 
    two basins of attraction. The other significant experimental observation relates to the fact that the sorted subpopulations relax back to the parental 
    distribution over a very slow time scale ( about two weeks ). It has been suggested that the attractor \cite{Brock,Huang2} is not 
    described by a simple valley with smooth ascending slopes in the potential landscape but corresponds to a valley with a highly rugged nature. 
    The valley has multiple `sub-attractors' in the ascending slopes of the basin around the central steady state. Cells are trapped in the transient states corresponding 
    to the sub-attractors. The outlier cells are these transiently stuck cells near the borders of the basin of attraction. 
    The outlier cells regenerate the parental distribution of Sca-1 levels through multiple steps of noise-driven transitions between 
    the metastable states thus slowing down the relaxation kinetics. The noise ( random fluctuations ) is a consequence of the stochastic 
    nature of gene expression. The experimentally observed broad dispersion of the parental histogram is the equilibrium distribution of the 
    occupancy of the sub-attractors by individual cells maintained via local noise-induced state transitions. Rugged landscapes with sub-attractors 
    are known to be a characteristic feature of complex, high-dimensional dynamical systems \cite{Brock,Huang2}. In reality, the cellular 
    state is described by the expression levels of a set of genes rather than a single gene, i.e., the state space is high-dimensional. In flow-cytometry 
    (FACS) experiments, however, the data describe the expression levels of a single gene which effectively implies the projection of the state 
    space onto a single axis. To reveal the additional dimensions, Chang {\it{et al}} \cite{Chang} carried out real-time PCR and found the evidence of correlated heterogeneity in 
    other proteins. For example, the GATA1 mRNA levels exhibit a 260-fold increase in the Sca-1$^{low}$ subpopulation of cells over the Sca-1$^{high}$ 
    fraction. Thus the heterogeneity is transcriptome-wide rather than involving the expression of a single gene. In the next section, we present an integrative explanation of 
    the slow relaxation kinetics and associated features based on our model calculations. 
    
    We end this section with a few comments on the stochastic dynamics of the model proposed by Wang {\it{et al}} \cite{Wang}. In the deterministic 
    case, a subcritical pitchfork bifurcation separates a region of tristability from a 
 \begin{figure}
 \centering
 \subfloat[\:\:\:\:\:\:( i )]{%
 \includegraphics[scale=0.2,width=0.22\textwidth]{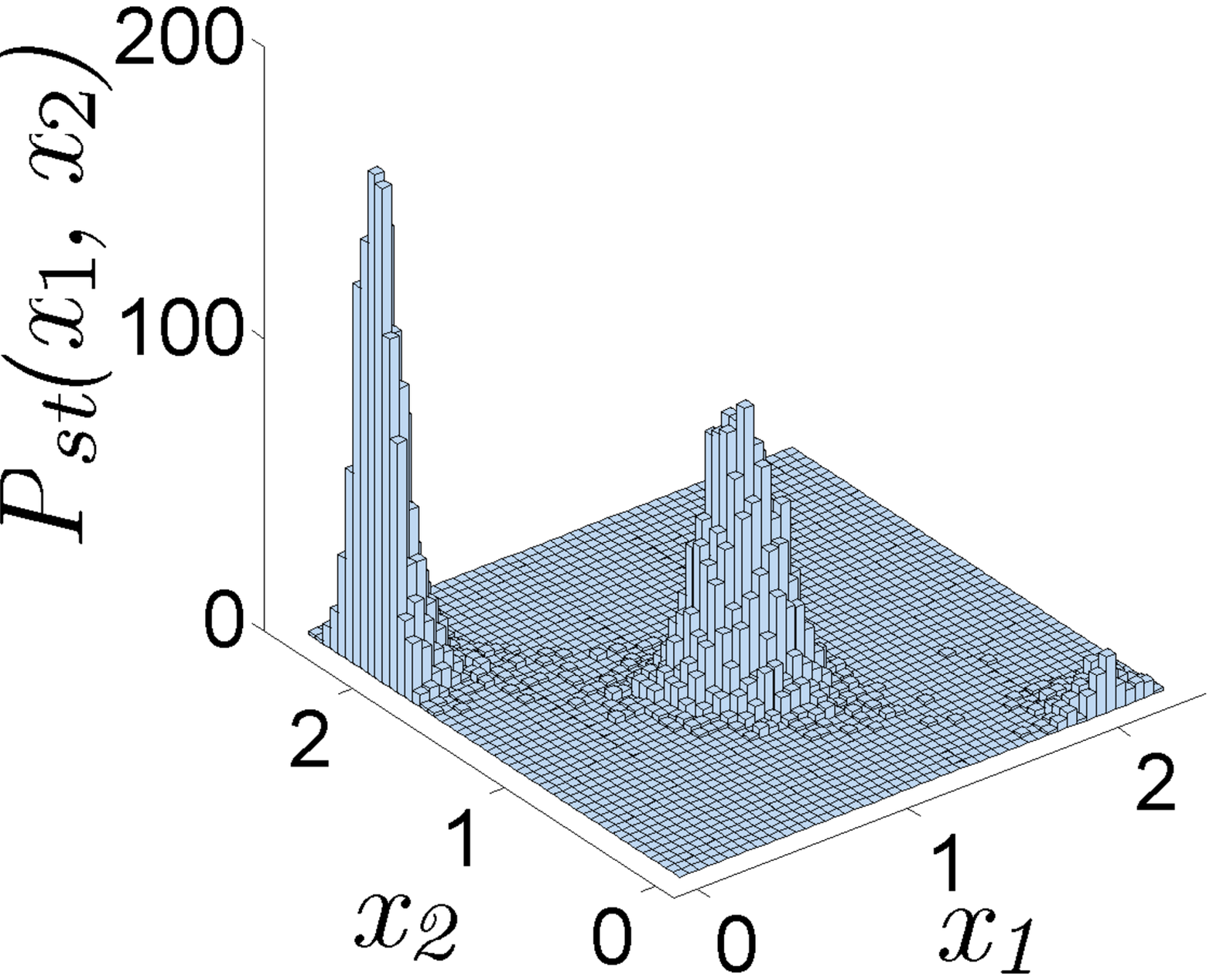}}
 \quad
 \subfloat[\:\:\:\:\:\:( ii )]{%
 \includegraphics[scale=0.2,width=0.22\textwidth]{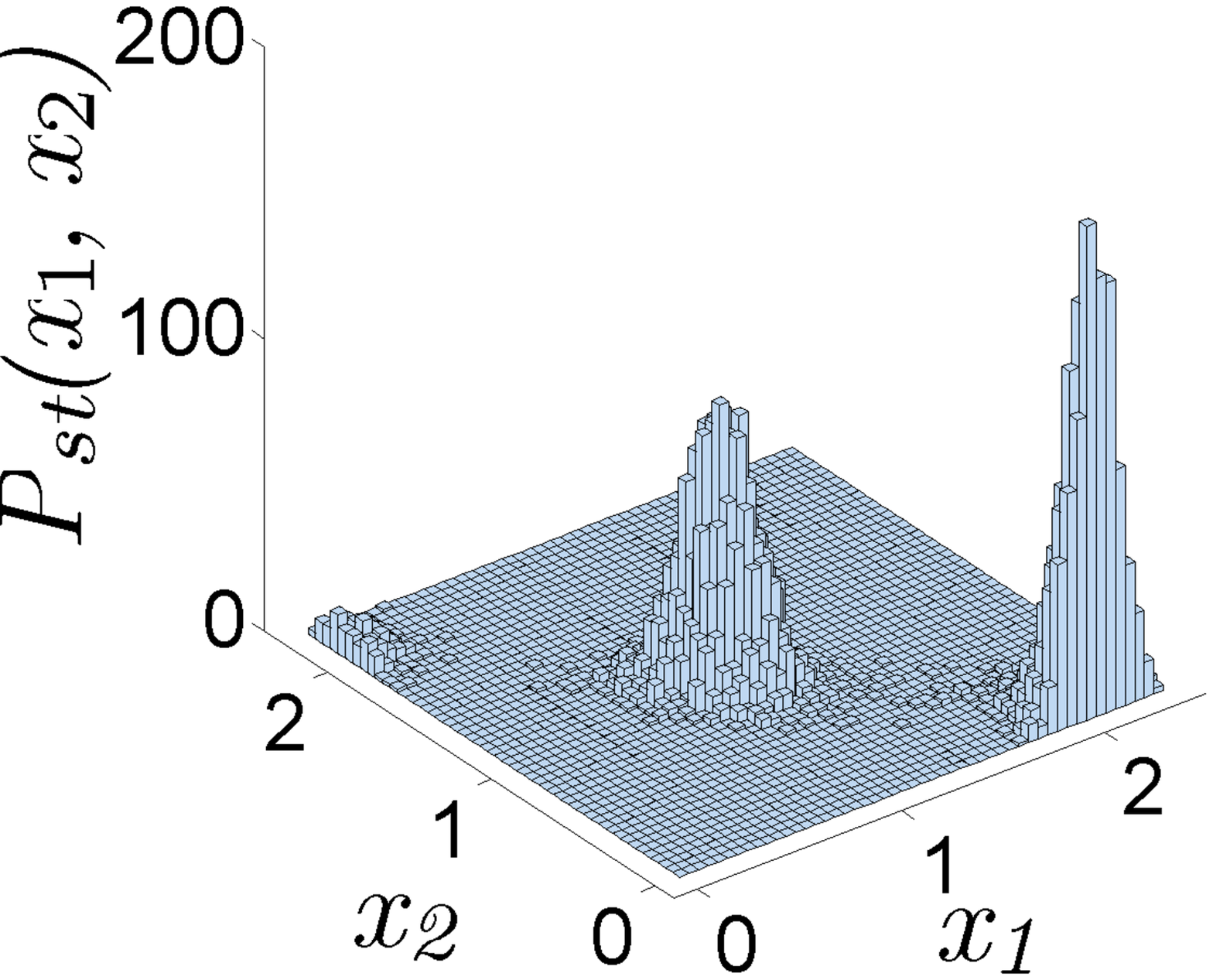}}
   
\subfloat[\:\:\:\:\:\:( iii )]{%
\includegraphics[scale=0.2,width=0.22\textwidth]{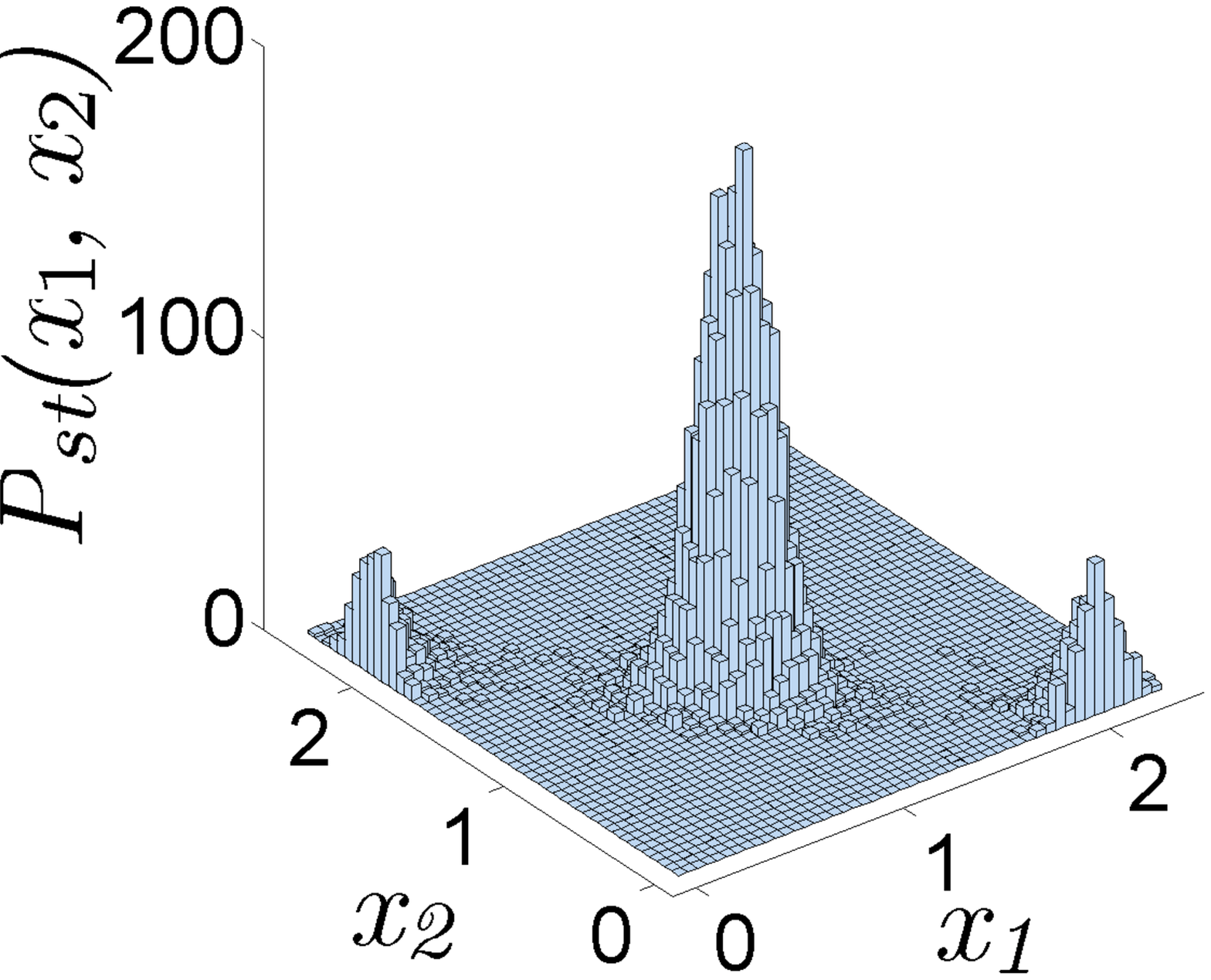}}

\caption{\small Steady state probability distributions, P$_{st}$({\it{ x{$_1$}}}, {\it{ x{$_2$}}}), obtained from the Langevin simulation of the model of Wang {\it{et al.}} \cite{Wang}. The simulation is carried out in the 
region of tristability. The probability distribution is trimodal with prominent peaks 
corresponding to ( i ) {\it x$_{2}$} $\gg$ {\it x$_{1}$}, ( ii ) {\it x$_{1}$} $\ll$ {\it x$_{2}$} and ( iii ) {\it x$_{1}$} $\approx$ {\it x$_{2}$}. The parameter values are as given in Table 1.
     \normalsize }
\end{figure}
    region of bistability. Figure 7 shows the results of the Langevin simulation in the region of tristability. 
    The parameter values used are mentioned in Table 1. The noise strength is fixed at $\Gamma$ = 0.15. The three stable steady states have three distinct 
    basins of attraction. The steady state probability distribution, P{$_{st}$}({\it{x{$_1$}}}, {\it{x{$_2$}}}), has the form determined by the location of the initial state, i.e., to which basin it belongs. The initial 
    state {\it{x{$_1$}}} = 0.01, {\it{x{$_2$}}} = 2.1 falls in the basin of attraction of the stable steady state corresponding 
    to {\it{x{$_1$}}}$\ll$ {\it{x{$_2$}}} (figure 7(i)). Figures 7(ii) and 7(iii) are obtained with the initial states {\it{x{$_1$}}} = 2.1, {\it{x{$_2$}}} = 0.01 and 
    {\it{x{$_1$}}} = 1.1, {\it{x{$_2$}}} = 1.1 respectively. In the last case, the subpopulation of undifferentiated cells is the most prominent one. In this model, cell differentiation 
    is possible without crossing a bifurcation point. By choosing the appropriate initial state, a cell may be made to adopt a specific lineage. 
    In the presence of noise, three distinct subpopulations can coexist: one undifferentiated and two differentiated ones. In the post-bifurcation 
    scenario, the undifferentiated subpopulation of cells no longer exists. In the region of tristability, a cell may be induced to differentiate 
    or change lineages by increasing/decreasing the amounts of appropriate TFs. This strategy possibly forms the basis of induced pluripotency and 
    lineage reprogramming \cite{MacArthur,Enver}. The trimodal probability distribution in the region of tristability is, however, in sharp contrast with the experimentally 
    observed single broad distribution of Sca-1 proteins in an ensemble of cells \cite{Chang}.
    
    \section*{4. Early signatures of bifurcation}
    
    A bifurcation involves a regime shift from one type of attractor dynamics to another. Some early signatures of the impending regime 
    shift are the critical slowing down and its associated effects, namely, a rising variance and the lag-1 autocorrelation function as the bifurcation (critical) 
    point is approached \cite{Scheffer,Scheffer1}. In the case of our model, the regime shift is from monostability to bistability at 
    the supercritical pitchfork bifurcation point. The model involves two variables {\it{x$_1$}} and {\it{x$_2$}} for which the stability of a steady state 
    is determined by the eigenvalues of the Jacobian matrix \textbf{J} for the dynamical system defined by 
    \begin{equation}
   \textbf{J} = 
    \left (
    \begin {array}{cc}
    
    \frac{\partial f(x{_1},x{_2})}{ \partial x{_1}} & \frac{\partial f(x{_1},x{_2})}{ \partial x{_2}}\\
    \frac{\partial g(x{_1},x{_2})}{ \partial x{_1}} & \frac{\partial g(x{_1},x{_2})}{ \partial x{_2}}\
    \end {array}
    \right )_{ss} 
   \end{equation}
   where the suffix `ss' denotes that the computation of the derivatives has to be carried out in the steady state. The functions 
   {\it{f}} ({\it{x{$_1$}}}, {\it{x{$_2$}}}) and {\it{g}} ({\it{x{$_1$}}}, {\it{x{$_2$}}}) are given by the right hand side expressions 
   in equations (1) and (2) respectively. A steady state is stable if the real parts of the eigenvalues, $\lambda{_i}$s ({\it{i}} = 1, 2), are negative. 
   Let $\lambda_{max}$ be the real part of the dominant eigenvalue of \textbf{J}, i.e., the one with the largest real part. One can define a return time T{$_R$} 
   which is an average measure of the time taken by a dynamical system to regain a stable steady state after being weakly perturbed from it. 
   The return time T{$_R$} is given by $T_{R}=\frac{1}{\mid\lambda_{max}\mid}$ \cite{ Strogatz,VanNes,Wissel}. The parameter $\lambda_{max}$ is a measure of the stability of a 
   state and becomes zero at the bifurcation point indicating a loss in the stability of the 
 \begin{figure}[h]
 \centering
 \subfloat[\:\:\:\:\:\:\:\:\:\:\:\:\:\:\:( i )]{%
 \includegraphics[scale=0.15,width=0.2\textwidth]{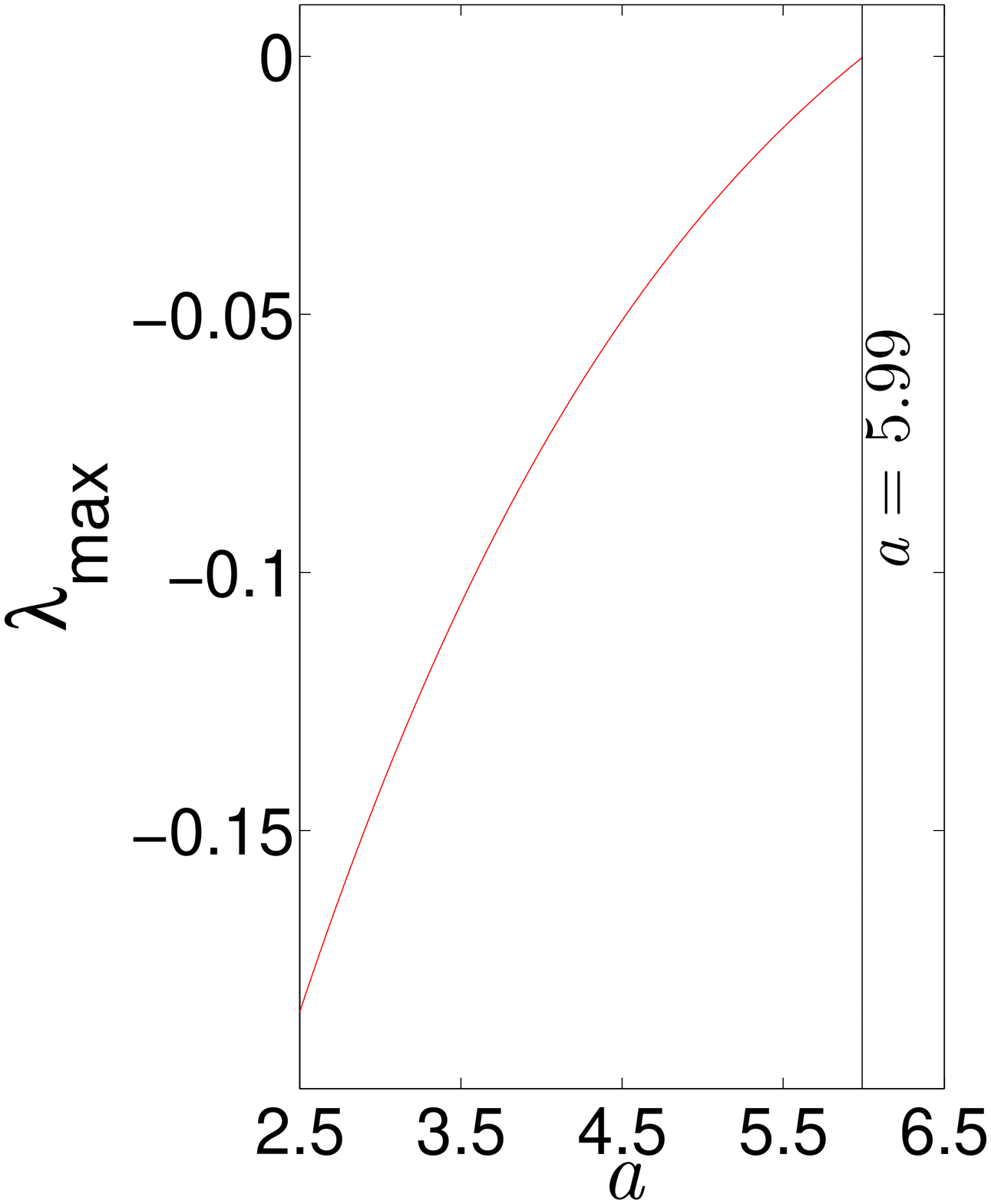}}
 \quad
 \subfloat[\:\:\:\:\:\:\:\:\:\:\:\:\:\:\:( ii )]{%
 \includegraphics[scale=0.15,width=0.2\textwidth]{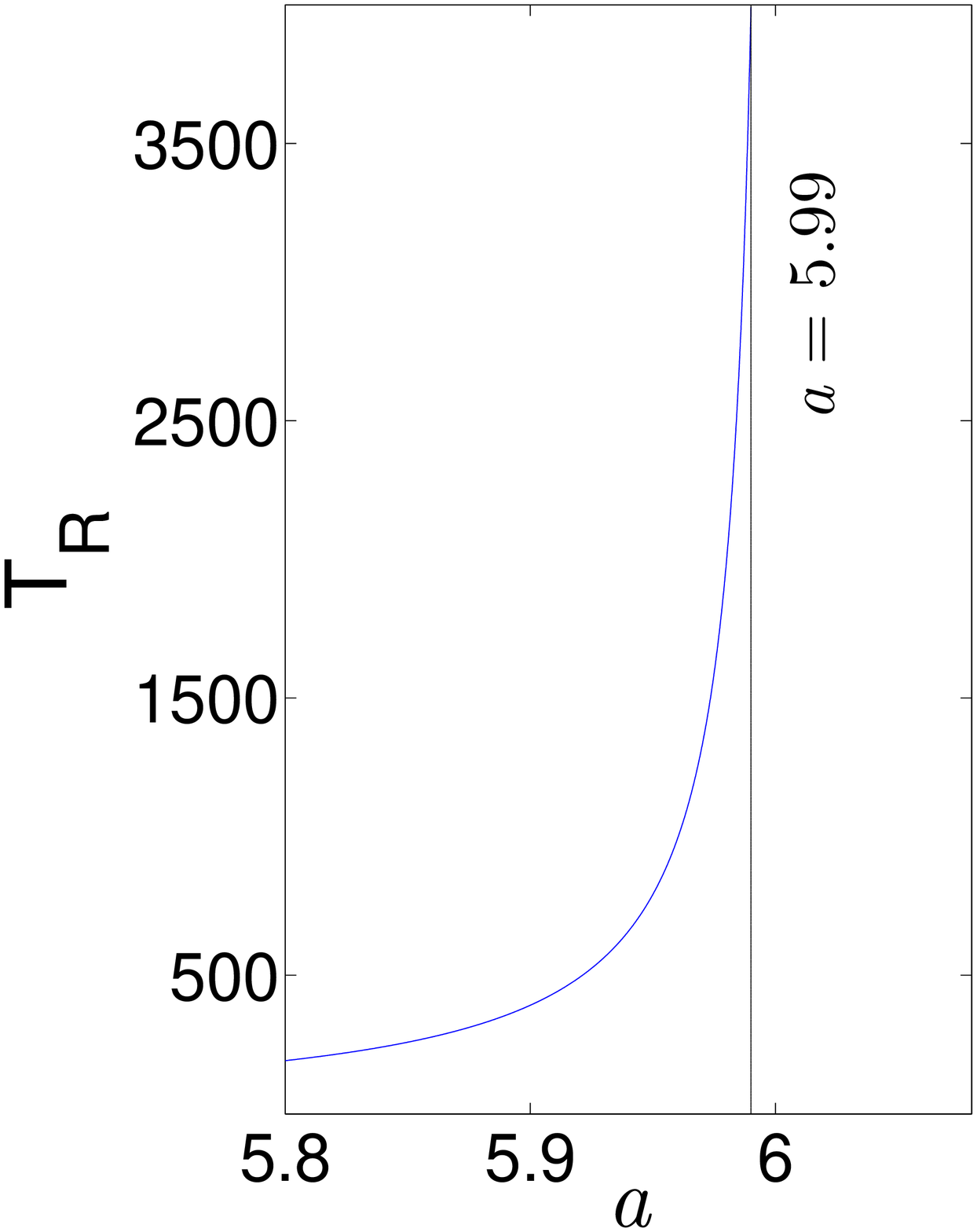}}
 
 \subfloat[\:\:\:\:\:\:\:\:\:\:\:\:\:\:\:( iii )]{%
 \includegraphics[scale=0.15,width=0.2\textwidth]{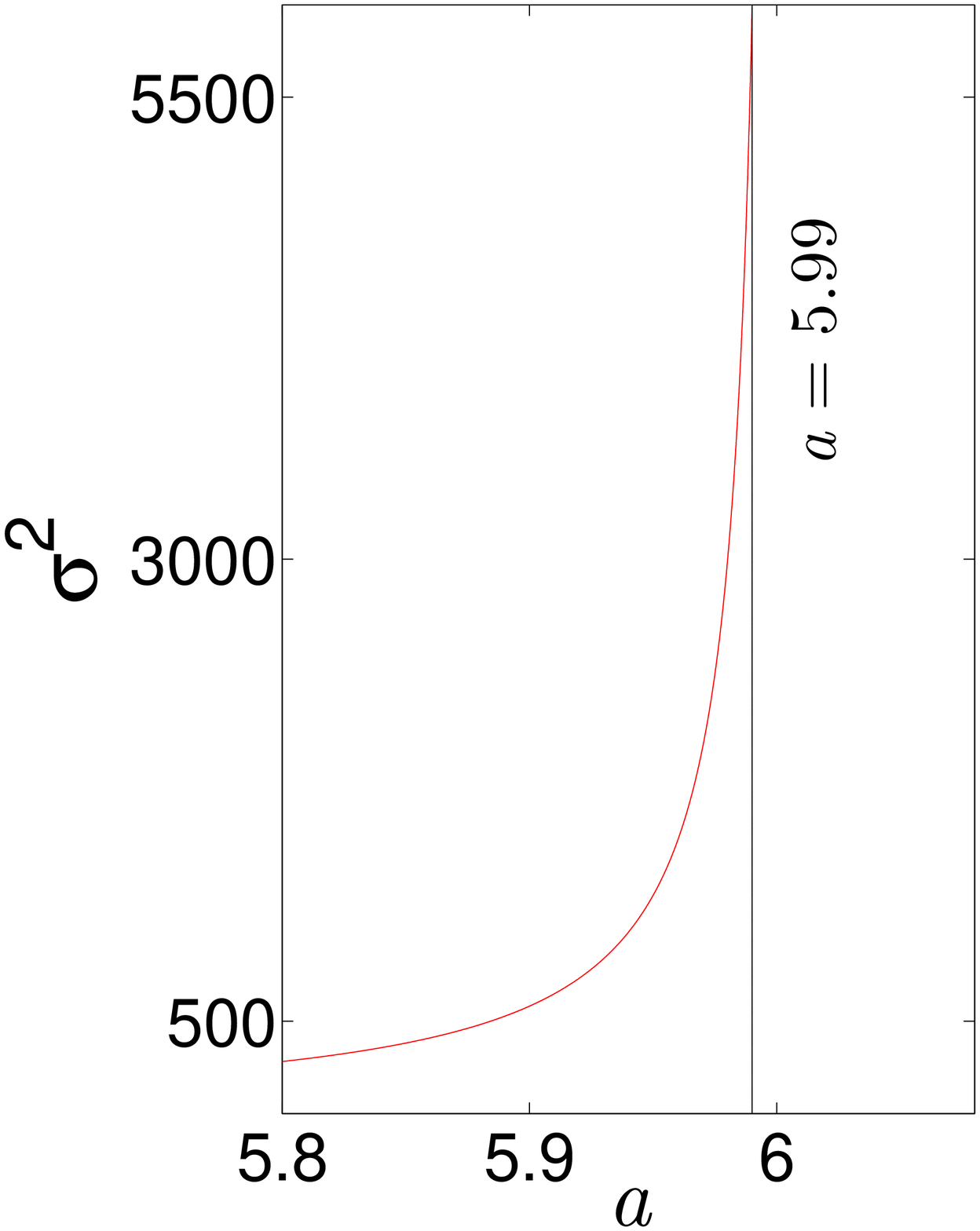}}
 \quad
 \subfloat[\:\:\:\:\:\:\:\:\:\:\:\:\:\:\:( iv )]{%
 \includegraphics[scale=0.15,width=0.19\textwidth]{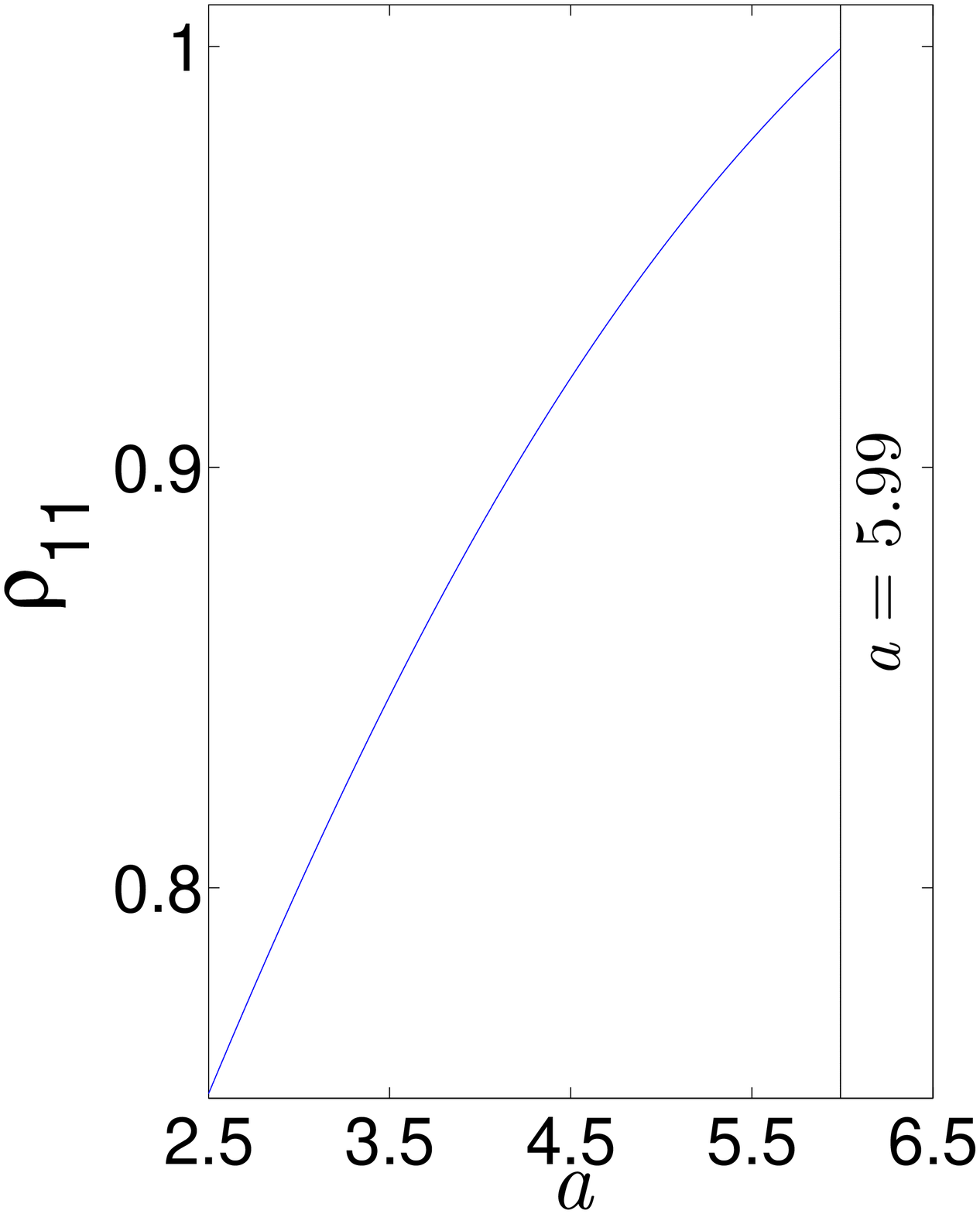}}
\caption{\small The variations of (i) $\lambda_{max}$, (ii) the return time T{$_R$}, (iii) the variance $\sigma^{2}$ and (iv) the 
lag-1 autocorrelation function $\rho_{11}$($\tau$ = 1) as a funtion of the bifurcation parameter {\it{a}}. The various quantities exhibit characteristic signatures 
as the bifurcation point {\it{a}} = 5.99 is approached. The parameter values are same as in the case of figure 2 (i) shown in Table 1.
\normalsize }
\end{figure}
   specific state. The return time T{$_R$} thus diverges as the bifurcation is 
   approached and the phenomenon is described as the critical slowing down. Experimental observations of such slowing down include the transition from the G2 growth phase to the mitotic 
   phase of the eukaryotic cell division cycle \cite{Sha}, a collapse transition at a 
   \begin{figure}
   \centering
 \includegraphics[scale=0.15,width=0.19\textwidth]{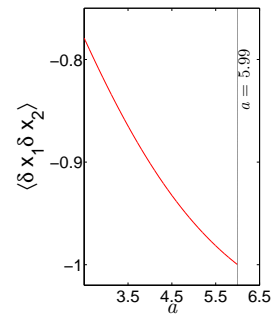} 
  \caption {\small The variation of the normalized covariance $\langle\delta{x{_1}}\:\:\delta{x{_2}}\rangle$ versus the bifurcation parameter {\it{a}}. 
   At the bifurcation point {\it{a}} = 5.99, the variables x{$_1$} and x{$_2$} are perfectely anticorrelated.
   \normalsize }
\end{figure}
   critical experimental condition in a laboratory population of budding yeast \cite{Dai} and a 
   similar light-induced regime shift in a population of cyanobacteria \cite{Veraart}. The critical slowing down close to a bifurcation point implies 
   that the system's intrinsic rates of change are lowered so that the state of the system at time {\it{t}} closely resembles the state of the system at time {\it{t}} - 1 ( time is discretized ). 
   This increased memory is measured by the lag-1 autocorrelation function. The magnitude of the function thus rises as the bifurcation point is 
   approached and becomes maximal at the point itself. The basin of attraction of a stable steady state becomes flatter as the bifurcation point at which the state loses its stability is approached. 
   Hence a given perturbation brings about a greater shift in the system's state variables, i.e., an increasing variance. 
   
   While the stability parameter $\lambda_{max}$ is determined from the Jacobian matrix, the computation of the variance and the lag-1 autocorrelation function 
   requires stochastic approaches. One such approach is based on the linear noise approximation, the basic methodology of which is explained in Refs. \cite{Van,Elf,Pal}. In the following, we describe 
   the major steps of the computational procedure. We consider N different chemical entities participating in R elementary reactions. The concentrations of the different chemical species 
   are represented in the form of the vector \textbf{x} = ({\it x${_1}$}, {\it x${_2}$},.........., {\it x${_N}$} )${^T}$ where T denotes the transpose. The state of the dynamical system is represented by \textbf{x} which changes due to the occurence of the chemical reactions. 
   The deterministic dynamics of the system are described by the rate equations
   \begin{equation}
    \frac{ dx{_i}}{dt}\ = \overset{R}{\underset{j = 1}{\sum}\:\:}S_{ij}\,f_{j}(\mathbf{x})\:\:(i=1,2,...,N)
    \label{arate}
   \end{equation}
   where S${_i}{_j}$, {\it {i}} = 1, 2, $\cdots$, {\it{ N}}, {\it {j}} = 1, 2,$\cdots$, {\it {R}} are the elements of a stoichiometric matrix \textbf{S}. The number of molecules of the chemical 
   component {\it {i}} changes from X{$_i$} to X{$_i$} + S{$_{ij}$} when the {\it{j}} th reaction takes place.
   
   \noindent 
   In a compact notation
    \begin{equation}
    \mathbf{\mathbf{\overset{.}{x}}=S f(x)},\:\:\mathbf{f}\left(\mathbf{x}\right)
     =(f_{1}(\mathbf{x}),\cdots,f_{R}(\mathbf{x}))^{T}
     \label{arate2}
      \end{equation}
      where \textbf{f(\textbf{x})} defines the reaction propensity vector. The steady state vector \textbf{x${_s}$} is determined from the condition 
      $\mathbf{\overset{.}{x}} = 0$, i.e., $\mathbf{f({x{_s}})} = 0$. Let $\delta\textbf{x}$ denote a weak perturbation applied to the steady state with 
      $\delta$\textbf{x} = \textbf{x} - \textbf{x{$_s$}}.    
       Retaining only terms linear in $\delta\textbf{x}$ one gets 
      
      \begin{equation}
       \frac{d(\delta\bf{x})}{dt} = \mathbf{J} \mathbf{\delta{x}}
       \label{arrte1}
      \end{equation}
      where \textbf{J} is the Jacobian matrix defined in equation (7). The elements of \textbf{J} are given by
      
    \begin{equation}
     J{_i}{_j} = \overset{R}{\underset{k=1}{\sum}\:\:} S{_i}{_k}\:\: \frac{\partial f{_k}} {\partial x{_j}}
     \label{arrte2}
    \end{equation}
    
    In the living cell, the biomolecules participating in the cellular reaction are frequently small in number so that a stochastic description of the dynamics is 
    more appropriate. The Chemical Master Equation describes the rate of change of the probability distribution 
    P(X${_1}$, X${_2}$,$\cdots$, X${_N}$, t) of the numbers of the different types of chemical species. The Chemical Master Equation is not exactly solvable in most cases and one has to take 
    recourse to various approximations in order to compute the relevant quantities. In the linear noise approximation, the steady state probability distribution ($\frac{dP} {dt}$) = 0 is given by a multivariate 
    Gaussian distribution. Also, the covariance of the fluctuations about the deterministic steady state is given by the fluctuation-dissipation (FD) relation 
    \begin{equation}
     \mathbf{J}\:\mathbf{C}  + (\mathbf{J}\:\mathbf{C}){^T}  + \mathbf{D} = 0
     \label{arrte3}
    \end{equation}
   where \textbf{J} is the Jacobian matrix, \textbf{C} = $\langle\delta\mathbf{x}\:\:\delta\mathbf{x{^T}}\rangle$
   is the covariance matrix, the diagonal elements of which are the variances, and \textbf{D} is the diffusion matrix. The matrix \textbf{D} has the form 
   \begin{equation}
    \mathbf{D} = \mathbf{S}\: diag \:( \mathbf{f}(\mathbf{x}) )\: \mathbf{S}{^T}
    \label{arrte4}
   \end{equation}
   where diag (\textbf{f(\textbf{x})}) is a diagonal matrix with the elements f${_j}(\textbf{x}),\:\:  j = 1, 2, \cdots, R$. With the knowledge of the \textbf{J} and \textbf{D} matrices, the elements of the covariance matrix, 
   C${_i}{_j}$=$\langle\delta{x{_i}}\:\:\delta{x{_j}}\rangle$ ({\it i}, {\it j} = 1, 2, $\cdots$, N), specially the variances, can be determined from the FD relation (12). The time correlation matrix for $\delta\textbf{x}$ is 
   given by 
   \begin {equation}
    \langle\delta\mathbf{x} (t+\tau) (\delta(\mathbf{x}(t)){^T}\rangle = exp \:(\mathbf{J} \tau)\: \mathbf{C}
    \label{arrte5}
   \end {equation}
   The diagonal elements of the matrix are the autocovariances and $\tau$ defines the lag time. With suitable normalization, the lag-1 autocorrelation function 
   for the {\it i} th chemical species is 
   \begin{equation}
   \rho{_i}{_i}=\frac{\langle\delta x_{i}(t+1)\delta x{}_{i}(t)\rangle}{\sqrt{\mbox{var}\left(x_{i}(t+1)\right)}
   \sqrt{\mbox{var}\left(x_{i}(t)\right)}}
   \label{autolag}
   \end{equation}
   The covariance C${_i}{_j}$ can be normalized  in a similar manner on dividing by the factor $\sqrt{\mbox{var}{x}{_i}}$ $\sqrt{\mbox{var}{x}{_j}}$.
   
   We now report on the results corresponding to the early signatures of an approaching bifurcation point in our model. 
   The deterministic rate equations in equations (1) and (2) describe the dynamics of the model. The stability parameter 
   $\lambda{_m}{_a}{_x}$ is computed as the eigenvalue of the Jacobian matrix in equation (7) with the largest real part. 
   Figures  8 (i) and (ii) show the variations of $\lambda{_m}{_a}{_x}$ and the return time {\it T}${_R}$ versus the bifurcation parameter 
   {\it a} with the other parameter values the same as in the case of Figure 5(i) (Table 1). One finds that the value of $\lambda{_m}{_a}{_x}$ tends to zero and the return 
   time {\it T}${_R}$ diverges as the bifurcation point {\it a} = 5.99 is approached. The steady state variance and the lag-1 autocorrelation function (equation (15)) 
   are computed using the FD relation in equation (12). The Jacobian matrix $\textbf{J}$ has the form shown in equation (11). The stoichiometric matrix is given by 
    \begin{equation}
     \textbf{S} = 
     \left (
    \begin {array}{ccccc}
     1   &  -1  &  -1  &   0  &   0 \\
     0   &   0  &  -1  &   1  &  -1 
     \end{array}
     \right )
      \end{equation}
 The first and the second rows of \textbf{S} correspond to the proteins X{$_1$} and X{$_2$} respectively. 
 The ``elementary composite reactions'' \cite{Van,Elf} considered for the 
 computations are obtained from equations (1) and (2):
  \begin{equation}
  \begin{array}{c} 
  X{_1}\xrightarrow{\;\;\frac{ax{_1}}{S+x{_1}}+\frac{bS}{S+x{_2}}\;\;\;}X{_1}+1\\
  X{_1}\xrightarrow{\;\;k x{_1}\;\;\;}X{_1}-1\\
  X{_1}, X{_2} \xrightarrow{\;\;g x{_1} x{_2}\;\;\;} X{_1}-1, X{_2}-1\\
  X{_2} \xrightarrow{\;\;\frac{ax{_2}}{S+x{_2}}+\frac{bS}{S+x{_1}}\;\;\;} X{_2}+1\\
  X{_2}\xrightarrow{\;\;k x{_2}\;\;\;} X{_2}-1
  \end{array}
  \label{comporeact}
 \end{equation}
 The reaction propensity vector (equation 9) is:
 \begin{equation}
  \textbf{f(x)} =
  \left(
  \begin {array}{c}
   \frac{a\:x{_1}}{S+x{_1}}\ + \frac{b\:S}{S+x{_2}}\\
   k\:x{_1}\\
   g\:x{_1}\:x{_2}\\
   \frac{a\:x{_2}}{S+x{_2}}\ + \frac{b\:S}{S+x{_1}}\\
   k\:x{_2}
  \end {array}
   \right)
 \end{equation}
The diffusion matrix \textbf{D} in the FD relation can be calculated using equation(13) and with the knowledge of the stoichiometric 
matrix \textbf{S} as well as reaction propensity vector \textbf{f(x)}. Substituting the computed \textbf{J} and \textbf{D} matrices in the 
FD relation, the variances and the covariances are determined. Similarly, from equation(15), the lag-1 autocorrelation function $\rho_{11}$($\tau$ = 1) is 
calculated. Figures 8 (iii) and (iv) exhibit the plots of the variance $\sigma^2$ = $\langle\delta x{_1}^2\rangle$ of the fluctuations in the 
X{$_1$} protein levels and the corresponding lag-1 autocorrelation function $\rho_{11}$($\tau$ = 1) versus the 
bifurcation parameter {\it{a}}. Figures 8 (iii) and (iv) clearly demonstrate that the variance diverges and the lag-1 autocorrelation function becomes maximal 
as the bifurcation point is approached. In the case of the supercritical pitchfork bifurcation, the early signatures of an approaching bifurcation are obtained when the 
bifurcation point is reached from both the lower and higher values of the bifurcation parameter. In the latter case, i.e., in the region of bistability, both 
the stable steady states lose stability at the bifurcation point. In the model of Wang {\it{et al}} \cite{Wang}, one goes from a region of tristability to that of bistability as the 
bifurcation parameter {\it{a}} is decreased. In this case, the early signatures are obtained only when the central progenitor state loses stability. The other two steady states are stable on both 
sides of the bifurcation point. 

The results obtained in this and the previous sections provide the basis for a new interpretation of the experimental observations of Chang {\it{et al}} \cite{Chang}. In 
the N-dimensional state space, the weak deviation $\delta$\textbf{x} of the state vector \textbf{x} from the stable steady state evolves according to equation (10). The Jacobian 
matrix \textbf{J} has N eigenvalues, $\lambda _{i}$'s ({\it{i}} = 1, 2, $\cdots$, N ), the real parts of which are negative for the stability of the state. This ensures that the steady state is 
regained in the long time limit, i.e., $\delta$\textbf{x}$\rightarrow$ 0. For the sake of simplicity we assume the eigenvalues to be real. The time - dependent solution $\delta\textbf{x}$ of equation (10) is 
given by 
\begin{equation}
 \delta\mathbf{x}(t) = \overset{N}{\underset{i = 1}{\sum}\:\:}C_{i}\, \mathbf{V{_{\it{i}}}}\:\: e^{\lambda{_i} t}
\end{equation}
where the $\lambda{_i}$'s and the \textbf{V{$_i$}}'s ({\it{i}} = 1, 2, $\cdots$, N) are the eigenvalues and the eigenvectors of the Jacobian matrix \textbf{J}. The coefficients, C{$_i$}'s, are 
determined from a knowledge of the initial value $\delta$\textbf{x}(0) of $\delta$\textbf{x}({\it{t}}). An eigenvector\textbf{ V{$_i$}} sets the direction in the N-dimensional state 
space along which $\delta$\textbf{x}({\it{t}}) = e$^{\lambda{_i} t}$ \textbf{V{$_i$}} \cite{Strogatz}.
The dominant eigenvalue of \textbf{J}, $\lambda_{max}$, is associated with the slowest relaxation mode. In the limit of large time, the trajectories in the state space approach the steady state tangent to the 
slow eigendirection. Close to a bifurcation (critical) point, $\lambda_{max}$ $\rightarrow$ 0 so that the relaxation to the steady state becomes significantly slow ( critical slowing down ). 
Near criticality, the direction of slow relaxation is also the direction of large fluctuations the distribution of which is non-Gaussian \cite{Krotov}. The experimental distribution of the Sca-1 levels 
exhibits the features of criticality as described above. It is distinguished by an unexpectedly slow relaxation kinetics and a large variance, also its shape deviates from the Gaussian. The slow mode describes the 
collective dynamics of N gene expression levels. In the flow-cytometry experiment \cite{Chang}, the distribution of only one of the levels in a population of cells is 
recorded. The sorted subpopulations, namely, the Sca-1 outliers have been shown to possess distinct transcriptomes. This reflects the broad heterogeneity of the whole transcriptome described in terms of N gene expression levels. In the 
computational study of our model, N=2 and the gene expression levels correspond to the GATA1 ({{\it x$_1$}}) and PU.1 ({{\it x$_2$}}) proteins. These protein levels are correlated with those of the Sca-1 \cite{Chang} so that the 
broad heterogeneity in the level distribution of the latter implies a similar heterogeneity in the distribution of the GATA1 and PU.1 levels. We have computed the steady state probability distribution, P{$_{st}$}({\it{x{$_1$}}}, {\it{x{$_2$}}}), in the undifferentiated regime close to the bifurcation point 
( figures 5(i) and 6(i)). The distribution exhibits considerable heterogeneity, consistent with experimental observations. The variance of the distribution is a measure of the heterogeneity in the distribution and a large variance indicates that the system is close to the bifurcation point. As one moves away from the bifurcation point, 
the steady state probability distribution becomes less broad, i.e., the variance decreases ( figure 8(iii) ). The experimentally observed slow relaxation kinetics have been demonstrated in the case of our model through the computation of the return time T{$_R$} ( figure 8(ii)). As the bifurcation point is approached, T{$_R$} increases indicating 
slower kinetics, i.e., the time needed to regain the stable steady state becomes longer. As already explained, $T_{R}=\frac{1}{\mid\lambda_{max}\mid}$ with $\lambda_{max}$ associated with the slowest relaxation mode. The slow eigendirection in the state space is given by the eigenvector of the Jacobian matrix {\textbf{J}} (equation (7)) corresponding to the eigenvalue 
$\lambda_{max}$. In the case of stochastic dynamics, since the description is in terms of probability distributions rather than individual states, the slowing down is in terms of a longer time needed by the system to regain the parental distribution starting with the probability distributions associated with the sorted subpopulations. In the case of our two-variable model, another signature of criticality lies in the perfect anti-correlation between the variables 
{\it{x{$_1$}}} and {\it{x{$_2$}}}. This is shown in figure 9, in which the normalized covariance approaches the value $-1$ close to the bifurcation point. The anti-correlation is in agreement with the experimental observation that the proteins GATA1 and PU.1 inhibit each other's expression. A quantitative measure of anti-correlation can be obtained from the simultaneous measurement of the protein levels as a function of time. The anti-correlation confers 
distinct identities on the differentiated states, x$_1$ $\gg$ x$_2$ and x$_1$ $\ll$ x$_2$ respectively. The quantitative signatures of the approach to a bifurcation point, namely, the critical slowing down, the 
rising variance and the lag-1 autocorrelation function are experimentally measureable \cite{Sha,Dai,Veraart,Weinberger,Kaufmann}. Experiments detecting the critical slowing down are difficult to carry out. 
The other quantities can be determined using single-cell methodologies like flow-cytometry and time-lapse fluorescense spectroscopy \cite{Weinberger,Kaufmann}. The requirement that a bifurcation point be crossed for the occurrence of 
cell differentiation can be checked through the measurements of quantities providing the early signatures of an approaching regime shift. Such signatures would be missing in the case of pure noise-driven transitions. Our proposal that 
the proximity of a bifurcation point confers the experimentally observed features of broad heterogeneity and slow relaxation kinetics on the cell population can be tested in the experimental measurements mentioned even if the actual crossing of a 
bifurcation point is not seen. The linear noise approximation, used in this section, suffers from certain limitations close to the 
bifurcation point. The approximation yields steady state distributions which are Gaussians and hence cannot capture the skewness in the distributions that develops close to criticality. 
The results reported in this section using the linear noise approximation are, however, generic in nature and reflect trends that are widely accepted \cite{Scheffer,Scheffer1}.
\section*{5. Summary and discussion}

In this paper, we have studied a simple model of cell differentiation based on the two-gene motif of opposing TFs (figure 1(i)). 
A large number of studies have already been carried out on this motif in different contexts. The importance of the motif specially arises from the 
fact that one of the first synthetic biology applications, namely, the construction of a bistable switch was based on 
this motif \cite{Gardiner}. Contrary to the genetic switches, no cooperativity in the regulation of the gene expression is 
assumed in the present model and an additional feature involving the formation of a heterodimer of the two TFs is taken into account. 
These assumptions are consistent with the experimental findings on blood cell differentiation \cite{Chickermane, Bokes, Liew}. 
The combination of molecular sequestration due to heterodimer formation and non-cooperative positive feedback controls the dynamics 
of our model bringing about a supercritical pitchfork bifurcation from monostability to bistability. Each of the parameters {\it{a}}, {\it{b}}, {\it{k}}, 
{\it{g}}, serves as a bifurcation parameter ( figure 2 ). The monostable state represents the undifferentiated cell state 
whereas the two stable steady states in the region of bistability describe the differentiated cell states. A synthetic genetic circuit 
\cite{Chen} has been designed to demonstrate bistability based on molecular sequestration and positive feedback. Another synthetic 
circuit has been constructed based on a bistable toggle switch and an intercellular signalling system to illustrate phenotypic 
diversification on a Waddington-type landscape with the initial cell density playing a crucial role in the synthetic phenotypic 
diversification \cite{Sekine}. These examples establish the utility of simple model systems in providing useful insights on the physical 
principles governing the operation of natural genetic circuits.

In the study of our model, we have addressed both the deterministic and stochastic aspects of cell differentiation. Stochasticity in the 
form of noisy gene expression is known to have varied degrees of influence on the deterministic dynamics depending on the specific 
nature of the cellular system or process \cite{Kaern,Raj}. One crucial role of noise is in binary cell-fate decisions \cite{Balazsi} 
including cell differentiation giving rise to phenotypic heterogeneity in the cell population in the form of two distinct subpopulations. 
This is seen in the experiments on the synthetic genetic circuits mentioned earlier \cite{Chen,Sekine} as well as in a host of natural and synthetic 
systems \cite{Kaern,Raj,Balazsi}. Recent theoretical studies analyse the stochastic dynamics in terms of stochastic bifurcations 
characterized by a qualitative change in the structure of the stationary probability distribution, e.g., the transition from an unimodal 
to a bimodal distribution \cite{Zakharova}. The experiments carried out by Chang {\it{ et al}}. \cite{Chang} provide a clear pointer to the 
role of stochasticity ( noise ) in cell differentiation. A very recent study \cite{Ahrends} combines computational modeling with experiments to show that noise in protein abundance acting on a network of more than six positive feedbacks brings about the 
differentiation of pre-adipocytes at low rates. Again, positive feedback and noise play important roles in this type of cell differentiation. 
The important experimental observations of Chang {\it{et al}}. \cite{Chang}, namely, the slow kinetics of the regeneration of the parental distribution 
of the Sca-1 proteins from the sorted subpopulations and the multilineage priming of the undifferentiated cell population have given rise to a number of plausible 
explanations which are still debated \cite{Huang2}. Based on the investigation of the stochastic dynamics of our model, we have provided an integrative physical 
explanation of the major experimental features. We have discussed that a natural explanation of the slow kinetics and the broad heterogeneity arises from the proximity 
of the bifurcation point (criticality). This view has not been put forward before in the context of cell differentiation. Our stochastic simulation further provides a 
physical understanding of the multilineage priming of the progenitor cell population. The set of states which constitute the undifferentiated distribution evolve into 
two subsets once the bifurcation point is crossed. A subset is distinguished by the location of the member states in the basin of attraction of a specific stable 
steady state. Noise has the effect of spreading out the set of initial states so that two distinct subpopulations of differentiated cells are obtained once external signals induce 
the crossing of the bifurcation point. In the case of models exhibiting a region of tristability in the phase diagram, the possibility of pure noise-induced cell differentiation 
exists but in this case one obtains a trimodal steady state probability distribution as shown in figure 7. The fraction of the cell population in a particular category, namely, undifferentiated, 
lineage 1 and lineage 2, depends on the initial states.

Our study further provides a framework for investigating whether cell differentiation involves an approaching bifurcation or not. A large number of studies have been carried out so far \cite{Scheffer,Scheffer1} on 
the early signatures of sudden regime shifts occurring at the bifurcation points of the saddle-node type. In this type of bifurcation, a region of bistability separates two 
regions of monostability and bistability is accompanied by hysteresis. This type of bifurcation is quite common in cell biology \cite{Pomerening} including cell differentiation 
\cite{Chickermane,Ahrends}. The first proposal to utilize the concept of early signatures in elucidating the nature of gene expression dynamics was made in Ref. \cite{Pal}. In 
the case of our model of cell differentiation, we have shown ( figures 8 and 9 ) how experimentally measurable quantities like the critical slowing down, the variance, the 
covariance and the lag-1 autocorrelation function provide concrete signatures of an approaching bifurcation point. For the experimental system of Chang {\it{et al}} \cite{Chang} 
such measurements would tell us how close one is to criticality. Our proposal to check the proximity/crossing of a bifurcation point in a cell differentiation process is novel but straightforward. 
The idea may be tested in a synthetic circuit governing cell differentiation or in specific cases where the suggested experimental 
measurements are possible. Some of the experimentally observed features in this paper are not unique to the population of mouse hematopoietic cells. Dynamic variability at the single cell level and multilineage priming of 
the progenitor cell population appear to be the hallmarks of other cell systems close to differentiation \cite{Mac,Abranches}. The results obtained in our study are thus expected to be of relevance in investigating the general problem of cell differentiation.

\subsection*{Acknowledgments}

MP acknowledges the support by UGC, India, vide sanction Lett. No. F.2-8/2002(SA-I) dated 23.11.2011. IB acknowledges the support by CSIR, India, vide sanction Lett. No. 
21(0956)/13-EMR-II dated 28.04.2014.
    
\small 
\doublespacing


\begin{thebibliography}{40}
\bibitem{Zhou}Zhou J X and Huang S 2011 Trends in Genetics \textbf{27}, 55
\bibitem{MacArthur}MacArthur B D , Ma'ayan A and Lemischka I R 2009 Nat. Rev. Mol. Cell Biol. \textbf{10}, 672
\bibitem{Enver}Enver T, Pera M, Peterson C and Andrews P W 2009 Cell
Stem Cell \textbf{4}, 387
\bibitem{Furusawa}Furusawa C and Kaneko K 2012
Science \textbf{338}, 215
\bibitem{Ferrel}Ferrell J E 2012 Current Biology \textbf{22} , R458
\bibitem{Huang}Huang S , Eichler G , Bar-Yam Y
and Ingber D 2005 Phys. Rev. Lett. \textbf{94},
128701
\bibitem{Heiniemei}Hein\"{a}niemi M et al.2013 Nature Methods \textbf{10}, 577
\bibitem{Huang1}Huang S, Guo Y-P, May G and Enver T 2007 Dev. Biol. \textbf{305}, 695
\bibitem{Wang}Wang J, Xu L, Wang E and Huang S 2010 Biophys. J. \textbf{99}, 29
\bibitem{Roeder}Roeder I and Glauche I 2006 J. Theoret. Biol. \textbf{241}, 852
\bibitem{Chickermane}Chickarmane V, Enver T and Peterson C 2009 PloS Comput. Biol. \textbf{5}, e1000268
\bibitem{Bokes}Bokes P, King J R and Loose M 2009 Math. Med. Biol.\textbf{26}, 117
\bibitem{Duff}Duff C, Smith-Miles K, Lopes L and Tian T 2012 J. Math. Biol. \textbf{64}, 449
\bibitem{Liew} Liew C W et al. 2006 J. Biol. Chem. \textbf{281}, 28296
\bibitem{Kaern}K\ae{}rn M, Elston T C, Blake WJ and Collins J J 2005 Nat. Rev. Genet. \textbf{6}, 451
\bibitem{Raj}Raj A and van Oudenaarden A 2008 Cell \textbf{135}, 216
\bibitem{Balazsi}Bal\'{a}zsi G, van Oudenaarden A and Collins J J 2011 Cell \textbf{144}, 910
\bibitem{Chang}Chang H H, Hemberg M, Barahona M, Ingber D E and Huang S 2008 Nature \textbf{453}, 544
\bibitem{Fox}Fox R F, Gatland I R. Roy R and Vemuri G 1988 Phys. Rev. A \textbf{38}, 5938
\bibitem{Scheffer}Scheffer M et al. 2009 Nature \textbf{461}, 53
\bibitem{Scheffer1}Scheffer M et al. 2012 Science \textbf{338}, 344
\bibitem{Pal}Pal M, Pal A K, Ghosh S and Bose I 2013 Phys. Biol.\textbf{10}, 036010
\bibitem{Ghosh} Ghosh S, Pal A K and Bose I 2013 Eur. Phys. J. E \textbf{36} : 123 
\bibitem{Strogatz}Strogatz S H 1994 \textit{Nonlinear dynamics and chaos - with applications to 
physics, biology, chemistry and engineering} (Addison-Wesley)
\bibitem{Roualt} Rouault  H and Hakim V 2012 Biophys. J. \textbf{102}, 417
\bibitem{Chen} Chen D  and Arkin A.P 2012 Mol. Syst. Biol. 8: \textbf{620}.
\bibitem{Buchler} Buchler N  E and Louis M 2008 J. Mol. Biol. \textbf{384}, 1106
\bibitem{Buchler1} Buchler N  E and Cross F.R. 2009 Mol. Syst. Biol. 5:\textbf{ 272}.
\bibitem{Graft} Graf T and Stadtfeld M. 2008 Cell Stem Cell \textbf{3}, 480
\bibitem{Brock} Brock A, Chang H and Huang S 2009 Nat. Rev. Genet. \textbf{10}, 336
\bibitem{Huang2} Huang S 2009 Development \textbf{136}, 3853
\bibitem{Van} van Kampen N G 1992  \textit{Stochastic Processes in Physics and Chemistry }(North Holland, Amsterdam)
\bibitem{Foster}Foster D V, Foster J G, Huang S, Kauffman S 2009 J. Theoret. Biol.\textbf{260}, 589
\bibitem{VanNes}van Nes E H and Scheffer M 2007 American Naturalist \textbf{169}, 738
\bibitem{Wissel}Wissel C 1989 Oecologia \textbf{65}, 101 
\bibitem {Sha} Sha W, Moore J, Chen K, Lassaletta A D, Yi C S, Tyson J J, Sible J C, 2003 Proc. Natl. Acad. Sci. USA 100 (3) 975 
\bibitem{Dai}Dai L, Vorselen D, Korolev K S and Gore J 2012 Science \textbf{336}, 1175 
\bibitem{Veraart}Veraart A J, Faassen E J, Dakos V, van Nes E H, L\"{u}ring M and Scheffer M 2012 Nature \textbf{481}, 357
\bibitem{Elf}Elf J and Ehrenberg M 2003 Genome Res. \textbf{13} (11) 2475
\bibitem{Krotov} Krotov D, Dubuis J O, Gregor T. and Bialek W  2014  Proc. Natl. Acad. Sci. USA 111(10) : 3638 
\bibitem{Weinberger}Weinberger L S, Dar R D and Simpson M L 2008 Nature Gent.\textbf{ 40}, 466
\bibitem{Kaufmann}Kaufmann B B and van Oudenaarden A 2007 Curr. Opin. Genet. Dev.\textbf{ 17}, 107
\bibitem{Gardiner} Gardner T S, Cantor C R and Collins JJ 2000 nature\textbf{ 403}, 339
\bibitem{Sekine} Sekine R {\it {et al}}  2011 Proc. Natl. Acad. Sci. USA \textbf{108}, 17969
\bibitem{Zakharova} Zakharova A,  Kurths J, Vadivasova T and Koseska A 2011 PloS ONE \textbf{6}, e 19696
\bibitem{Ahrends} Ahrends R, Ota A, Kovary K M, Kudo T, Park B O and Teruel M N 2014 Science \textbf{344}, 1384
\bibitem{Pomerening}Pomerening J R 2008 Curr. Opin. Biotechnol. \textbf{19} (4), 381
\bibitem{Mac} MacArthur B D and Lemischka I R 2013 Cell \textbf{154}, 484
\bibitem{Abranches} Abranches E {\it{et al}}. 2014 Development 141, 2770


\end{thebibliography}
\end{document}